\documentclass[fleqn,12pt]{article}
\usepackage{amssymb,graphics,epsfig,here}

\def\beq   {\begin{equation}}
\def\eeq   {\end{equation}}
\def\beqd  {\begin{displaymath}}
\def\eeqd  {\end{displaymath}}
\def\beqaa {\begin{eqnarray}}
\def\eeqaa {\end{eqnarray}}

\def\sz{\ifmmode{\tilde{\chi}^0} \else{$\tilde{\chi}^0$} \fi}
\def\sw{\ifmmode{\tilde{\chi}} \else{$\tilde{\chi}$} \fi}

\newcommand{\be}[1]{\begin{equation} \label{(#1)}}
\newcommand{\ee}{\end{equation}}
\newcommand{\baq}[1]{\begin{eqnarray} \label{(#1)}}
\newcommand{\eaq}{\end{eqnarray}}

\newcommand{\ba}{\begin{array}}
\newcommand{\ea}{\end{array}}

\begin{document}
\pagestyle{empty}

\begin{flushright}
  DCPT-04-30 \\
  IPPP-04-15 \\
  UWThPh-2004-11 \\
  WUE-ITP-2004-012 \\
  hep-ph/0406190
\end{flushright}

\vfill

\begin{center}

{\Large {\bf
A T-odd asymmetry in neutralino production and decay
}}\\

\vspace{10mm}

{\large
A.~Bartl$^a$, H.~Fraas$^b$, S.~Hesselbach$^a$,
K.~Hohenwarter-Sodek$^a$, G.~Moortgat-Pick$^c$
}

\vspace{6mm}

$^a${\it Institut f\"ur Theoretische Physik, Universit\"at Wien, A-1090
Vienna, Austria}\\
$^b${\it Institut f\"ur Theoretische Physik und Astrophysik,
Universit\"at W\"urzburg,}\\
{\it Am Hubland, D-97074 W\"urzburg, Germany}\\
$^c${\it IPPP, University of Durham, Durham DH1 3LE, U.K.}

\end{center}

\vfill

\begin{abstract}

We study CP-violating effects in neutralino
production and subsequent decay within the Minimal
Supersymmetric Standard Model with complex parameters
$M_1$ and $\mu$. The observable we propose is a T-odd asymmetry based
on a triple product in neutralino production
$e^+e^-\rightarrow\tilde{\chi}^0_i\tilde{\chi}^0_2$, $i=1,\ldots,4\,$,
with subsequent leptonic three-body decay
$\tilde{\chi}^0_2\rightarrow\tilde{\chi}^0_1\ell^+\ell^-$, 
$\ell=e,\mu$, at an $e^+e^-$ linear
collider with $\sqrt{s} = 500$~GeV and polarised beams.
We provide compact analytical formulae for the cross section and 
the T-odd asymmetry
taking into account the complete spin correlations between
production and decay.
We give numerical predictions for the cross section
and the T-odd asymmetry. The asymmetry can go up to 10\,\%.

\end{abstract}

\vfill

\newpage
\pagestyle{plain}


\section{Introduction}

In the Standard Model (SM) the phase in
the Cabibbo-Kobayashi-Maskawa (CKM) matrix is the only
source of CP violation.
The small amount of CP violation in the SM, however,
is not sufficient to explain the baryon-antibaryon
asymmetry of the universe \cite{bau}.
The Lagrangian of the Minimal Supersymmetric Standard Model (MSSM)
contains several complex parameters, which can
give rise to new CP-violating phenomena \cite{mssmcpv}.
In the neutralino sector of the MSSM two
complex parameters appear, which lead to
CP-violating effects in neutralino production
and decay. These parameters are the $U(1)$ gaugino 
mass parameter $M_1$ and the higgsino 
mass parameter $\mu$. The phase of the $SU(2)$ 
gaugino mass parameter $M_2$ can be eliminated 
by an $U(1)$-symmetry of the model.

One of the main goals of a future $e^+e^-$ linear
collider will be a careful study of the properties
of supersymmetric (SUSY) particles \cite{B1}.
The neutralinos $\tilde{\chi}^0_i$, $i=1,\cdots,4\,$,
will be particularly interesting, because the lightest 
neutralino $\tilde{\chi}^0_1$ is expected to be the
lightest SUSY particle (LSP), which is stable if $R$-parity is conserved.
The second lightest neutralino,
$\tilde{\chi}^0_2$, will presumably be among the lightest
visible SUSY particles. Therefore, the study of 
production and decay of the neutralinos $\tilde{\chi}^0_i$
\cite{B1, bfm, Gounaris:2002pj} 
and a precise determination of the underlying supersymmetric
parameters $M_1$, $M_2$, $\mu$ and $\tan\beta$
including the phases $\phi_{M_1}$ and $\phi_\mu$ of 
$M_1$ and $\mu$ will play an important role at future linear colliders. 
The phases of $M_1$ and $\mu$ cause CP-violating effects already
at tree level. Therefore these effects could be large and thus be
measurable at a high luminosity $e^+e^-$ linear collider.
Methods to determine these parameters
in neutralino and chargino production
have been presented in
\cite{kneur, Barger:1999tn, choichargino, Choi:2001ww}.
In particular the method of \cite{Choi:2001ww}
is based on the analysis of the masses and production cross sections
of only the light neutralinos $\tilde{\chi}^0_1$ and
$\tilde{\chi}^0_2$ and the light chargino $\tilde{\chi}^\pm_1$
and will allow the determination of the parameters
$|M_1|$, $M_2$, $|\mu|$ and the phases $\phi_{M_1}$ and $\phi_\mu$.
All these methods involve only CP-even 
quantities, in which the signs of the phases $\phi_{M_1}$ 
and $\phi_\mu$ can not be determined.
For an unambiguous determination of $\phi_{M_1}$ 
and $\phi_\mu$ including their signs 
one has to rely on CP-sensitive observables.

The phases of the complex parameters are constrained or 
correlated by the experimental upper limits  on the electric
dipole moments of electron, neutron and the atoms 
${}^{199}$Hg and ${}^{205}$Tl \cite{edmexp}.
In a constrained MSSM the restrictions on the phases can be rather 
severe. However, there 
may be cancellations between the contributions of different 
complex parameters, which allow larger values for the phases
\cite{edm}.
For example, in a constrained MSSM and if substantial
cancellations are present, $\phi_\mu$ is restricted 
to $|\phi_\mu| \lesssim 0.1\pi$, whereas $\phi_{M_1}$
and $\phi_A$, the phase of the trilinear scalar coupling
parameter, turn out to be essentially unconstrained,
but correlated with $\phi_\mu$ \cite{Choi:2004rf}.
Moreover, the restrictions are very model dependent. 
For example, when also lepton flavour violating terms 
are included, then the restriction on $\phi_\mu$ 
may disappear \cite{Bartl:2003ju}.
Therefore it is necessary to determine the phases in an unambiguous way
by measurements of CP-odd observables.

A useful tool to study these CP-violating effects are
T-odd observables, based on triple products
of momenta or spin vectors of the particles involved
\cite{Choi:1999cc, tripleproducts}. In this paper we study
a T-odd asymmetry in neutralino production
\begin{equation}\label{process1}
e^+e^-\rightarrow \tilde{\chi}^0_k\tilde{\chi}^0_2, \qquad
k=1,\cdots,4\,,
\end{equation}
with subsequent leptonic three-body decay
\begin{equation}\label{process2}
\tilde{\chi}^0_2\rightarrow \tilde{\chi}^0_1\ell^+\ell^-,
\end{equation}
with $\ell=e,\mu$. Using the triple product of 
the initial electron momentum $\vec{p}_{e^-}$ and
the two final lepton momenta $\vec{p}_{\ell^+}$ and $\vec{p}_{\ell^-}$,
\begin{equation} \label{tpdef}
O_T=\vec{p}_{\ell^+}\cdot (\vec{p}_{\ell^-}\times\vec{p}_{e^-}),
\end{equation} 
we define a T-odd asymmetry
\begin{equation} \label{Asy}
A_T = \frac{\int\mathrm{sign}\{
 O_T\}
 |T|^2d\mbox{lips}}{{\int}|T|^2d\mbox{lips}},
\end{equation}
where ${\int}|T|^2d\mbox{lips}$ is proportional to 
the cross section 
$\sigma(e^+e^- \to \tilde{\chi}^0_k\tilde{\chi}^0_2
\to \tilde{\chi}^0_k\tilde{\chi}^0_1 \ell^+ \ell^-)$. 
It is essential to include the spin correlations between production
and decay, because otherwise the numerator and hence the
asymmetry $A_T$ would vanish.
This observable has the advantage that it is not
necessary to reconstruct the momentum of the decaying
neutralino. The asymmetry $A_T$, Eq.~(\ref{Asy}), is odd
under the naive time-reversal operation. By CPT it is
a CP-sensitive asymmetry, if final-state interactions
and finite-widths effects are unimportant. We will neglect
these effects, because they are of higher order. By its
definition, the asymmetry $A_T$ is the difference of the 
number of events with the final lepton $\ell^+$ above and
below the plane spanned by $\vec{p}_{\ell^-}\times \vec{p}_{e^-}$,
normalised by the sum of these events.

In this paper we will first present the complete analytic
formulae at tree level for the cross section and for the 
asymmetry (\ref{Asy}) of the processes
(\ref{process1}) and (\ref{process2}), taking
into account the full spin correlations between
production and decay of the neutralino $\tilde{\chi}^0_2$
\cite{Moortgat-Pick:1999di}.
We will study the dependence of the asymmetry and the cross section
on the phases $\phi_\mu$ and $\phi_{M_1}$ and
present detailed numerical results for an $e^+ e^-$ linear collider
with $\sqrt{s}=500$~GeV and polarised beams.
Analogous CP asymmetries in neutralino production
and subsequent two-body decays have been studied
in \cite{Bartl:2003tr}.
CP asymmetries based on triple products in decays of scalar fermions
have been discussed in \cite{Bartl:2003ck}.
CP-odd observables involving 
the polarisation of the outgoing $\tau$ leptons from neutralinos
decaying via two-body decays
$\tilde{\chi}^0_2 \to \tilde{\tau}_1^\pm \tau^\mp \to 
  \tilde{\chi}^0_1 \tau^\pm \tau^\mp$
have been analysed in \cite{Ataupol}.
A systematic study of the impact of the supersymmetric phases on
neutralino, chargino and selectron production at linear colliders has
been performed in \cite{Choi:2004rf}.
A Monte Carlo study of the asymmetry (\ref{Asy}) for 
$e^+e^-\to \tilde{\chi}^0_1\tilde{\chi}^0_2\to
\tilde{\chi}^0_1\tilde{\chi}^0_1 \ell^+\ell^-$, $l=e,\mu$,
including SM backgrounds and detector effects
has been given in \cite{Aguilar-Saavedra:2004dz},
however, no analytic formulae for the
cross section and the asymmetries have been given there. 
In \cite{sf3papers} the impact of the phases 
$\phi_{A_\tau}$, $\phi_{A_t}$, $\phi_{A_b}$, $\phi_\mu$ and
$\phi_{M_1}$
on the two-body decays of the third generation sfermions has been
analysed in detail.
The influence of these phases on the polarisation of the outgoing
fermions in third generation sfermion decays has been studied in
\cite{sfermionpol}.

In Section 2 we shortly present the basics of the spin density matrix
formalism. In section 3 we give the analytic expressions for the
asymmetry $A_T$ and the cross sections. Section 4 contains numerical
results and discussions. In section 5 we present our
conclusions. Appendices A and B contain some details of the formalism
and of the analytic expressions.

\section{Formalism}

\subsection{Lagrangian and couplings}

\begin{figure}[ht!]
\hspace{-2.1cm}
\begin{minipage}[t]{6cm}
\begin{center}
{\setlength{\unitlength}{0.6cm}
\begin{picture}(5,9)
\put(-3,-10){\includegraphics{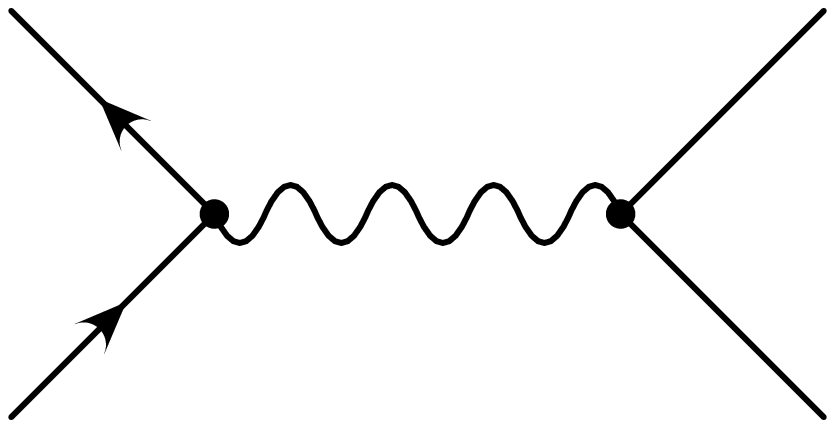}}
\put(1.2,0.4){{\small $e^{-}(p_1)$}}
\put(10.8,0.4){{\small $\tilde{\chi}^0_j(p_4)$}}
\put(1.2,4.9){{\small $e^{+}(p_2)$}}
\put(10.8,4.9){{\small $\tilde{\chi}^0_i(p_3)$}}
\put(6.5,3.3){{\small $Z^0$}}
\end{picture}}
\end{center}
\end{minipage}
\hspace{-0.7cm}
\begin{minipage}[t]{5cm}
\begin{center}
{\setlength{\unitlength}{0.6cm}
\begin{picture}(2.5,5)
\put(-4,-10){\includegraphics{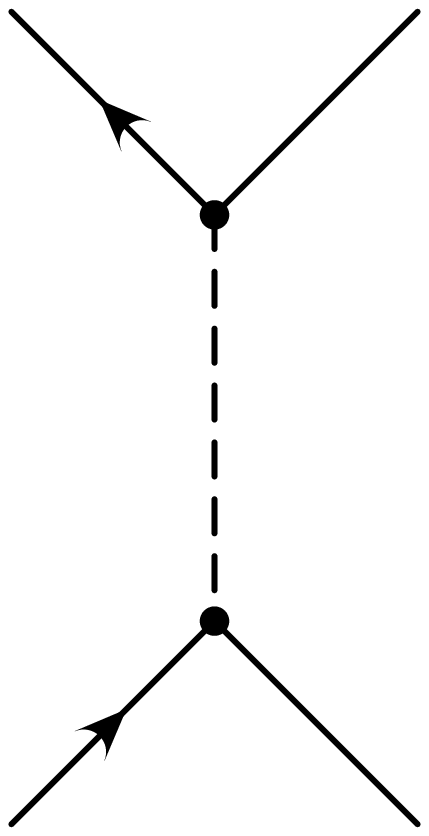}}
\put(2.1,-1.5){{\small $e^{-}(p_1)$}}
\put(2.1,6.8){{\small $e^{+}(p_2)$}}
\put(7.8,-1.5){{\small $\tilde{\chi}^0_j(p_4)$}}
\put(7.8,6.8){{\small $\tilde{\chi}^0_i(p_3)$}}
\put(4.4,2.8){{\small $\tilde{e}_{L,R}$}}
 \end{picture}}
\end{center}
\end{minipage}
\begin{minipage}[t]{5cm}
\begin{center}
{\setlength{\unitlength}{0.6cm}
\begin{picture}(2.5,5)
\put(-4.5,-10){\includegraphics{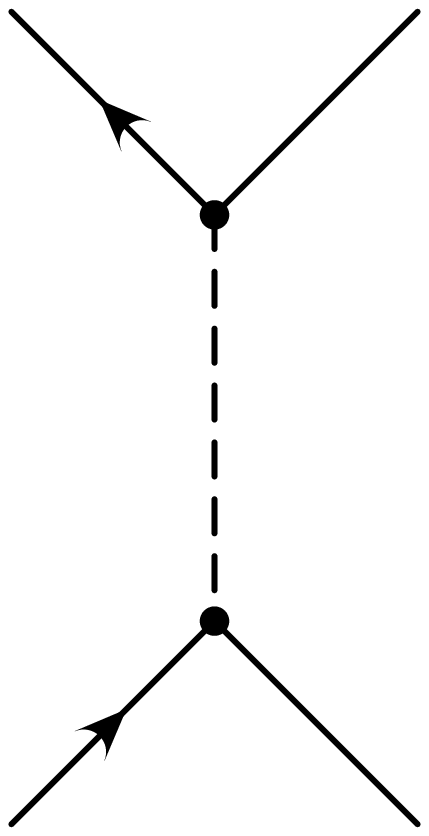}}
\put(1.7,-1.5){{\small $e^{-}(p_1)$}}
\put(1.7,6.8){{\small $e^{+}(p_2)$}}
\put(7.4,-1.5){{\small $\tilde{\chi}^0_i (p_3)$}}
\put(7.4,6.8){{\small $\tilde{\chi}^0_j (p_4)$}}
\put(3.9,2.8){{\small $\tilde{e}_{L,R}$}}
\end{picture}}
\end{center}
\end{minipage}
\vspace{1cm}
\caption{\label{Fig:FeynProd}Feynman diagrams of the production process
  $e^{+}e^{-}\to\tilde{\chi}^0_i\tilde{\chi}^0_j$.}
\vspace{1cm}
\end{figure}

\begin{figure}[ht!]
\hspace{-0.5cm}
\begin{minipage}[t]{5cm}
\begin{center}
{\setlength{\unitlength}{0.8cm}
\begin{picture}(5,7)
\put(-5.2,-5.2){\includegraphics{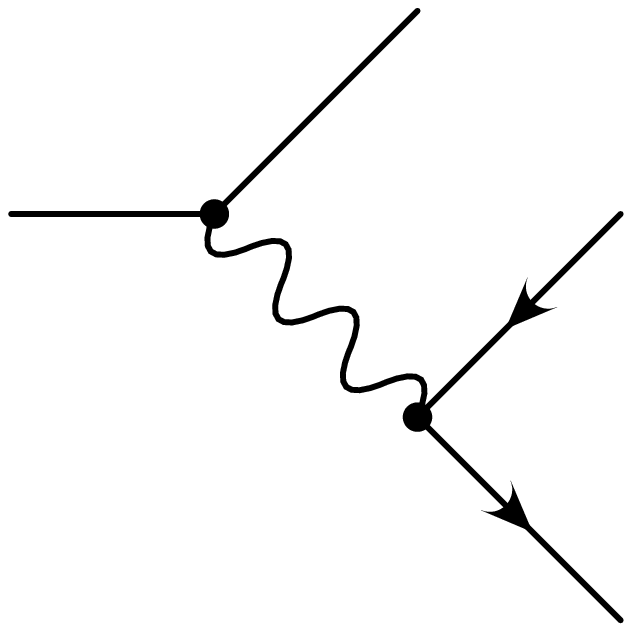}}
\put(4.7,1.7){{\small $\ell^{-}(p_7)$}}
\put(0.3,5.5){{\small $\tilde{\chi}^0_{i}(p_3)$}}
\put(4.7,5.5){{\small $\ell^{+}(p_6)$}}
\put(3.7,7.1){{\small $\tilde{\chi}^0_{k}(p_5)$}}
\put(2.2,3.8){{\small $Z^0$}}
\end{picture}}
\end{center}
\end{minipage}
\hspace{0.cm}
\begin{minipage}[t]{5cm}
\begin{center}
{\setlength{\unitlength}{0.8cm}
\begin{picture}(5,7)
\put(-5.2,-6){\includegraphics{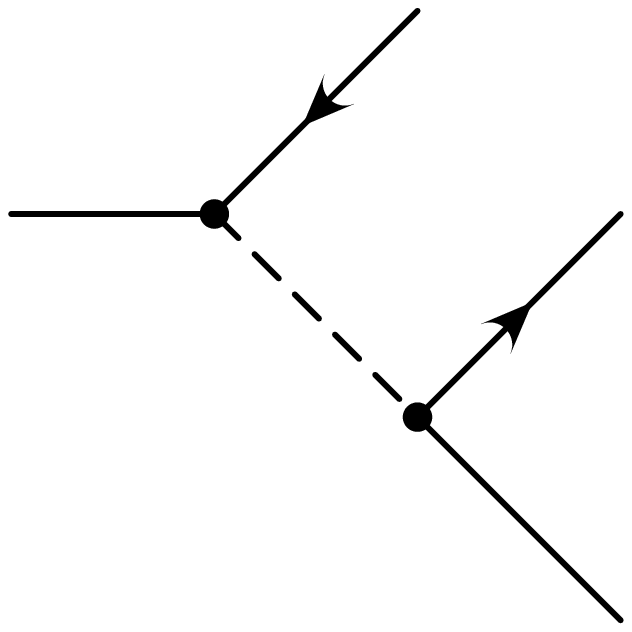}}
\put(4.7,5.5){{\small $\ell^{-}(p_7)$}}
\put(3.7,7.1){{\small $\ell^{+}(p_6)$}}
\put(4.7,1.5){{\small $\tilde{\chi}^0_{k}(p_5)$}}
\put(0.3,5.5){{\small $\tilde{\chi}^0_{i}(p_3)$}}
\put(2.1,3.9){{\small $\tilde{\ell}_{L,R}$}}
\end{picture}}
\end{center}
\end{minipage}
\begin{minipage}[t]{5cm}
\begin{center}
{\setlength{\unitlength}{0.8cm}
\begin{picture}(5,7)
\put(-5.2,-6){\includegraphics{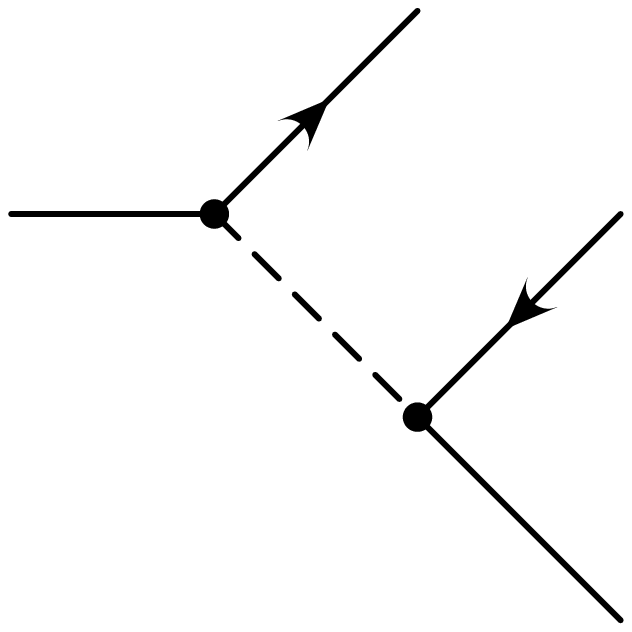}}
\put(3.7,7.1){{\small $\ell^{-}(p_7)$}}
\put(4.7,5.5){{\small $\ell^{+}(p_6)$}}
\put(0.3,5.5){{\small $\tilde{\chi}^0_{i}(p_3)$}}
\put(4.7,1.5){{\small $\tilde{\chi}^0_{k}(p_5)$}}
\put(2.1,3.9){{\small $\tilde{\ell}_{L,R}$}}
\end{picture}}
\end{center}
\end{minipage}
\vspace{-1.2cm}
\caption{\label{Fig:FeynDecay}Feynman diagrams of the three-body decay
$\tilde{\chi}^0_i\to \tilde{\chi}^0_k\ell^{+}\ell^{-}$.}
\end{figure}

The production process 
$e^+e^-\rightarrow \tilde{\chi}^0_i\tilde{\chi}^0_2$,
$i=1,\cdots,4\,$, proceeds via  
$Z^0$ and $\tilde{e}_{L,R}$ exchange
(Fig.~\ref{Fig:FeynProd}). In the
decay process 
$\tilde{\chi}^0_2\rightarrow \tilde{\chi}^0_1\ell^+\ell^-$, $\ell=e,\mu\,$,
$Z^0$ and $\tilde{\ell}_{L,R}$ 
exchanges contribute (Fig.~\ref{Fig:FeynDecay}).

The interaction Lagrangians for these processes are \cite{haberkane}
(in our notation and conventions we follow \cite{Moortgat-Pick:1999di})
\begin{equation}
\mathcal{L}_{Z^0\ell^+\ell^-} =
  -\frac{g}{\cos\Theta_W}Z_{\mu}\bar{\ell}\gamma^{\mu}
  [L_\ell P_L+R_\ell P_R]\ell, 
\end{equation}
\begin{equation}
\mathcal{L}_{Z^0\tilde{\chi}^0_m\tilde{\chi}^0_n} =
 \frac{1}{2}\frac{g}{\cos\Theta_W}Z_{\mu}{\bar{\tilde{\chi}}}^0_m\gamma^{\mu}
 [O''^L_{mn}P_L+O''^R_{mn}P_R]{\tilde{\chi}}^0_n,
\end{equation}
\begin{equation}
\mathcal{L}_{\ell\tilde{\ell}\tilde{\chi}^0_k} =
 g f^L_{\ell k} \bar{\ell} P_R \tilde{\chi}^0_k \tilde{\ell}_L +
 g f^R_{\ell k} \bar{\ell} P_L \tilde{\chi}^0_k \tilde{\ell}_R + 
 \textrm{h.c.},  
\end{equation}
with $m,n,k=1,\cdots,4\,$ and the couplings
\begin{equation}
f^L_{\ell k} = 
 -\sqrt{2}[\frac{1}{\cos\Theta_W}(T_{3\ell}-e_\ell\sin^2\Theta_W)N_{k2}
 + e_\ell\sin\Theta_WN_{k1}]\,,
\end{equation}
\begin{equation}
f^R_{\ell k} = -\sqrt{2}e_\ell\sin\Theta_W[\tan\Theta_WN^*_{k2}-N^*_{k1}]\,,
\end{equation}
\begin{equation}
O''^L_{mn}=-\frac{1}{2}(N_{m3}N^*_{n3}-N_{m4}N^*_{n4})\cos2\beta-\frac{1}{2}(N_{m3}N^*_{n4}+N_{m4}N^*_{n3})\sin2\beta,
\end{equation}
\begin{equation}
O''^R_{mn}=-O''^{L*}_{mn},
\end{equation}
\begin{equation}
L_\ell=T_{3\ell}-e_\ell\sin^2\Theta_W,\quad R_\ell=-e_\ell\sin^2\Theta_W ,
\end{equation}
where $P_{L,R}=\frac{1}{2}(1\mp\gamma_5)$,
$g$ is the weak coupling constant
($g=e/\cos\Theta_W$, $e>0$). $e_\ell$ and 
$T_{3\ell}$ are the charge (in units of $e$) and 
the third component of the
weak isospin of the fermion $\ell$. $\Theta_W$ is the
weak mixing angle and $\tan\beta=v_2/v_1$
is the ratio of the
vacuum expectation values of the Higgs fields.
The unitary $(4\times4)$-matrix $N_{mk}$
which diagonalises the complex symmetric 
neutralino mass matrix is given in the basis
$(\tilde{\gamma}, \tilde{Z}, \tilde{H}^0_a, \tilde{H}^0_b)$
\cite{bfm}.

\subsection{Amplitudes}

From these interaction Lagrangians the helicity amplitudes 
$T_P^{\lambda_i\lambda_j}(\alpha)$ and $T_{D,\lambda_i}(\alpha)$
for production and decay can be calculated.
Here $\alpha$ denotes the production or decay channel
and the respective exchanged particle.
$\lambda_i$ and  $\lambda_j$ are the helicities of the neutralinos
$\tilde{\chi}^0_i$ and $\tilde{\chi}^0_j$.
In $T_{D,\lambda_i}(\alpha)$
we suppress the label of the helicity
of the final neutralino $\tilde{\chi}^0_k$.
The amplitudes $T_P^{\lambda_i\lambda_j}(\alpha)$ for the production
process
$e^{-}(p_1) e^{+}(p_2)\to \tilde{\chi}^0_i(p_3)\tilde{\chi}^0_j(p_4)$
(see Fig.~\ref{Fig:FeynProd}) are
\begin{equation}
T_P^{\lambda_i\lambda_j} = T_P^{\lambda_i\lambda_j}(s,Z) +
  T_P^{\lambda_i\lambda_j}(t,\tilde{e}_{L}) +
  T_P^{\lambda_i\lambda_j}(t,\tilde{e}_{R}) +
  T_P^{\lambda_i\lambda_j}(u,\tilde{e}_L) +
  T_P^{\lambda_i\lambda_j}(u,\tilde{e}_R),
\end{equation}
where
\begin{eqnarray}
& &\mbox{\hspace*{-.4cm}}
T_P^{\lambda_i\lambda_j}(s,Z)
=\frac{g^2}{\cos^2\Theta_W}
 \Delta^s(Z)\bar{v}(p_2)\gamma^{\mu}(L_{\ell} P_L+R_{\ell}
P_R)u(p_1)\nonumber\\[-0.4cm]
& &\mbox{\hspace*{-.4cm}}
\phantom{T_P^{\lambda_i\lambda_j}(s,Z)=\frac{g^2}{\cos^2\Theta_W}\Delta^s(Z)}
\bar{u}(p_4,\lambda_j)\gamma_{\mu}
(O^{''L}_{ji} P_L+O^{''R}_{ji} P_R)v(p_3,\lambda_i),\label{eq_a1}\\[-0.2cm]
& &\mbox{\hspace*{-.4cm}}
T_P^{\lambda_i\lambda_j}(t,\tilde{e}_{L})
=-g^2 f^L_{e i}f^{L*}_{e j}
\Delta^t(\tilde{e}_L)\bar{v}(p_2) P_R v(p_3,\lambda_i)
\bar{u}(p_4,\lambda_j)P_L u(p_1),\label{eq_a2a}\\
& &\mbox{\hspace*{-.4cm}}
T_P^{\lambda_i\lambda_j}(t,\tilde{e}_{R})
=-g^2 f^R_{e i}f^{R*}_{e j}
\Delta^t(\tilde{e}_R)\bar{v}(p_2) P_L v(p_3,\lambda_i)
\bar{u}(p_4,\lambda_j)P_R u(p_1),\label{eq_a2}\\
& &\mbox{\hspace*{-.4cm}}
T_P^{\lambda_i\lambda_j}(u,\tilde{e}_L)
=g^2 f^{L*}_{e i} f^L_{e j}
\Delta^u(\tilde{e}_L)\bar{v}(p_2) P_R v(p_4,\lambda_j)
\bar{u}(p_3,\lambda_i)P_L u(p_1),\label{eq_a3a}\\
& &\mbox{\hspace*{-.4cm}}
T_P^{\lambda_i\lambda_j}(u,\tilde{e}_R)
=g^2 f^{R*}_{e i} f^R_{e j}
\Delta^u(\tilde{e}_R)\bar{v}(p_2) P_L v(p_4,\lambda_j)
\bar{u}(p_3,\lambda_i)P_R u(p_1)  \label{eq_a3}
\end{eqnarray}
with
\begin{equation} \label{eq_Zprop}
\Delta^s(Z) = \frac{i}{s - m_Z^2 + i m_Z\Gamma_Z},
\end{equation}
\begin{equation} \label{eq_selprop}
\Delta^t(\tilde{e}_{L,R}) = \frac{i}{t -
  m_{\tilde{e}_{L,R}}^2 + i m_{\tilde{e}_{L,R}}\Gamma_{\tilde{e}_{L,R}}},
\quad
\Delta^u(\tilde{e}_{L,R}) = \frac{i}{u -
  m_{\tilde{e}_{L,R}}^2 + i m_{\tilde{e}_{L,R}}\Gamma_{\tilde{e}_{L,R}}}
\end{equation}
and the Mandelstam variables
\begin{equation}
s = (p_1 + p_2)^2, \quad t = (p_1 - p_4)^2, \quad u = (p_1 - p_3)^2.
\end{equation}
$\Delta^s(Z)$, $m_Z$, $\Gamma_Z$,
$\Delta^{t,u}(\tilde{e}_{L,R})$, $m_{\tilde{e}_{L,R}}$,
$\Gamma_{\tilde{e}_{L,R}}$ denote the corresponding propagator, mass
and width of the exchanged particle.
The amplitudes  $T_{D,\lambda_i}(\alpha)$ for the decay
$\tilde{\chi}^0_i(p_3)\to\tilde{\chi}^0_k(p_5)\ell^{+}(p_6)\ell^{-}(p_7)$
(see Fig.~\ref{Fig:FeynDecay}) read
\begin{equation}
T_{D, \lambda_i} = T_{D, \lambda_i}(s_i,Z) +
  T_{D, \lambda_i}(t_i,\tilde{\ell}_L) + 
  T_{D, \lambda_i}(t_i,\tilde{\ell}_R) +
  T_{D, \lambda_i}(u_i,\tilde{\ell}_L) +
  T_{D, \lambda_i}(u_i,\tilde{\ell}_R)
\end{equation}
with
\begin{eqnarray}
& &\mbox{\hspace*{-.4cm}}
T_{D, \lambda_i}(s_i,Z)
=-\frac{g^2}{\cos^2\Theta_W}
\Delta^{s_i}(Z) \bar{u}(p_7) \gamma^{\mu} (L_{\ell} P_L+R_{\ell} P_R) v(p_6)
\nonumber\\[-0.4cm]
& &\mbox{\hspace*{-.4cm}}
\phantom{T_{D, \lambda_i}(s_i,Z)=-\frac{g^2}{\cos^2\Theta_W}\Delta^{s_i}(Z)}
\bar{u}(p_5) \gamma_{\mu} (O^{''L}_{ki} P_L+O^{''R}_{ki} P_R)
 u(p_3, \lambda_i), \label{eq_a4}\\[-0.2cm]
& &\mbox{\hspace*{-.4cm}}
T_{D, \lambda_i}(t_i,\tilde{\ell}_L)
=- g^2 f^L_{\ell k} f^{L*}_{\ell i}
\Delta^{t_i}(\tilde{\ell}_L)\bar{u}(p_7) P_R v(p_5)
\bar{v}(p_3, \lambda_i) P_L v(p_6),\label{eq_a5a}\\
& &\mbox{\hspace*{-.4cm}}
T_{D, \lambda_i}(t_i,\tilde{\ell}_R)
=-g^2 f^R_{\ell k} f^{R*}_{\ell i}
\Delta^{t_i}(\tilde{\ell}_R)\bar{u}(p_7) P_L v(p_5)
\bar{v}(p_3, \lambda_i) P_R v(p_6), \label{eq_a5}\\
& &\mbox{\hspace*{-.4cm}}
T_{D, \lambda_i}(u_i,\tilde{\ell}_L)
=+g^2 f^L_{\ell i} f^{L*}_{\ell k}
\Delta^{u_i}(\tilde{\ell}_L)\bar{u}(p_7) P_R u(p_3, \lambda_i)
\bar{u}(p_5) P_L v(p_6),\label{eq_a6a}\\
& &\mbox{\hspace*{-.4cm}}
T_{D, \lambda_i}(u_i,\tilde{\ell}_R)
=+g^2 f^R_{\ell i} f^{R*}_{\ell k}
\Delta^{u_i}(\tilde{\ell}_R)\bar{u}(p_7) P_L u(p_3, \lambda_i)
\bar{u}(p_5) P_R v(p_6)  \label{eq_a6}
\end{eqnarray}
and the kinematic variables
\begin{equation}
s_i = (p_6 + p_7)^2, \quad t_i = (p_3 - p_6)^2, \quad u_i = (p_3 - p_7)^2.
\end{equation}
The amplitude for the whole process
$e^+e^-\to \tilde{\chi}^0_i\tilde{\chi}^0_j\to
\tilde{\chi}^0_1\ell^+\ell^-\tilde{\chi}^0_j$,
with $\tilde{\chi}^0_i$ decaying, can be written
as
\begin{equation}
T=\Delta(\tilde{\chi}^0_i)\sum_{\lambda_i}
T_P^{\lambda_i\lambda_j}T_{D,\lambda_i},
\end{equation}
where $\Delta(\tilde{\chi}^0_i)=1/[p_3^2-m^2_i+im_i\Gamma_i]$,
$m_i$, $\Gamma_i$
are the propagator, mass and width of the 
decaying neutralino $\tilde{\chi}^0_i$.

\subsection{Cross section}

Following the formalism of \cite{Moortgat-Pick:1999di,Haber:1994pe},
the amplitude squared $|T|^2$ of the
combined processes of production 
$e^+e^-\to \tilde{\chi}^0_i\tilde{\chi}^0_j$ and decay 
$\tilde{\chi}^0_i\to\tilde{\chi}^0_1\ell^+\ell^-$
of the neutralino $\tilde{\chi}^0_i$, with
neutralino $\tilde{\chi}^0_j$ being unobserved, 
can be written as
\begin{eqnarray}
|T|^2 = 4|\Delta(\tilde{\chi}^0_i)|^2  
 \left\{ P(\tilde{\chi}^0_i\tilde{\chi}^0_j) D(\tilde{\chi}^0_i)
             +
            \sum^3_{a=1}\Sigma^a_P(\tilde{\chi}^0_i)
                    \Sigma^a_D(\tilde{\chi}^0_i)\right\},
\label{Tsquared}
\end{eqnarray}
where $a=1,2,3$ refers to the polarisation state 
of the neutralino $\tilde{\chi}^0_i$,
which is described by the polarisation vectors
$s^a(\tilde{\chi}^0_i)$ given in Appendix A.
$P(\tilde{\chi}^0_i\tilde{\chi}^0_j)$ and
$D(\tilde{\chi}^0_{i})$ are the terms 
of production and decay independent of the polarisation of the
decaying neutralino, whereas
$\Sigma^a_P(\tilde{\chi}^0_{i})$ and
$\Sigma^a_D(\tilde{\chi}^0_{i})$ are the terms 
containing the spin correlations between production and decay. 
According to our choice of the polarisation
vectors $s^a(\tilde{\chi}^0_i)$, Eqs.~(\ref{eq_21}) -- (\ref{eq_23}),
$\Sigma^3_P/P$ is the longitudinal polarisation,
$\Sigma^1_P/P$ is the transverse polarisation in the
production plane and $\Sigma^2_P/P$ is the
polarisation perpendicular to the production plane
of the neutralino $\tilde{\chi}^0_i$.
The explicit expressions for these contributions are given in Appendix B.
The complete expressions for the amplitude squared 
including spin-spin correlations
for the case when both neutralinos are decaying and for
longitudinally polarised
$e^{\pm}$ beams are given in \cite{Moortgat-Pick:1999di}.
The differential cross section in the laboratory
system is
\begin{equation}
d\sigma =
 \frac{1}{8E^2_b}|T|^2 
 (2\pi)^4 \delta^4(p_1 + p_2 - \sum_{i=4}^7 p_i)d 
  \mbox{lips}(p_3 \cdots p_{7}), 
\end{equation}
where $E_b$ is the beam energy and
$d\mbox{lips}(p_3\cdots p_{7})$ is the
Lorentz invariant phase space element.
Integrating over all angles gives the cross section for the combined
process of production and decay
\begin{eqnarray}
\sigma & = & \sigma(e^+e^- \to \tilde{\chi}^0_i\tilde{\chi}^0_j)
\cdot BR(\tilde{\chi}^0_i \to \tilde{\chi}^0_1 \ell^+ \ell^-) \nonumber\\
 & = & \frac{1}{2 E^2_b}
 \int \! P(\tilde{\chi}^0_i\tilde{\chi}^0_j) D(\tilde{\chi}^0_i)
 |\Delta(\tilde{\chi}^0_i)|^2
 (2\pi)^4 \delta^4(p_1 + p_2 - \sum_{i=4}^7 p_i)d 
  \mbox{lips}(p_3 \cdots p_{7}).
\label{crosss}
\end{eqnarray}

\section{Triple product and CP asymmetry}

In this section we will derive the analytic
formulae for the cross section
$\sigma(e^+e^- \to \tilde{\chi}^0_i\tilde{\chi}^0_j
\to \tilde{\chi}^0_j\tilde{\chi}^0_1 \ell^+ \ell^-)$
and for the asymmetry $A_T$, Eq.~(\ref{Asy}).
As can be seen from the numerator of $A_T$,
for this purpose we have to identify those
terms in $|T|^2$, Eq.~(\ref{Tsquared}), which contain
a triple product of the form Eq.~(\ref{tpdef}).
Triple products follow from expressions 
$i\epsilon_{\mu\nu\rho\sigma} k^\mu p^\nu q^\rho s^\sigma$
in the terms $\Sigma^a_P(\tilde{\chi}^0_{i})$ and
$\Sigma^a_D(\tilde{\chi}^0_{i})$,
where $k$, $p$, $q$, $s$ are 4-momenta and spins of the particles involved
(see Appendix B, Eqs.~(\ref{eq_sumsp1}) -- (\ref{eq_kinsp4})
and (\ref{eq_dssum}) -- (\ref{eq_dssub6})).
The expressions $i\epsilon_{\mu\nu\rho\sigma} k^\mu p^\nu q^\rho s^\sigma$ 
are imaginary and when multiplied by
the imaginary parts of the respective couplings they yield the terms
which contribute to the numerator of $A_T$, Eq.~(\ref{Asy}). 
Hence only the second term of Eq.~(\ref{Tsquared}), 
which is due to the spin correlations, contributes to the 
numerator of the asymmetry $A_T$.
It is convenient to split
$\Sigma^a_P(\tilde{\chi}^0_i)$ and $\Sigma^a_D(\tilde{\chi}^0_i)$
into T-odd terms
$\Sigma^{a,\mathrm{O}}_{P}(\tilde{\chi}^0_i)$ and 
$\Sigma^{a,\mathrm{O}}_{D}(\tilde{\chi}^0_i)$ containing the
respective triple product, and T-even terms
$\Sigma^{a,\mathrm{E}}_{P}(\tilde{\chi}^0_i)$ and
$\Sigma^{a,\mathrm{E}}_{D}(\tilde{\chi}^0_i)$
without triple products:
\begin{equation}
 \Sigma^a_P(\tilde{\chi}^0_i) =
  \Sigma^{a,\mathrm{O}}_P(\tilde{\chi}^0_i) +
  \Sigma^{a,\mathrm{E}}_P(\tilde{\chi}^0_i), \qquad
 \Sigma^a_D(\tilde{\chi}^0_i) =
  \Sigma^{a,\mathrm{O}}_D(\tilde{\chi}^0_i) +
  \Sigma^{a,\mathrm{E}}_D(\tilde{\chi}^0_i).
\end{equation}
Then the terms of $|T|^2$, Eq.~(\ref{Tsquared}), which
contribute to the numerator of $A_T$ are
\begin{equation} \label{CPinT}
 |T|^2 \supset 4|\Delta(\tilde{\chi}^0_i)|^2 
  \sum^3_{a=1} \left[\Sigma^{a,\mathrm{O}}_P(\tilde{\chi}^0_i)
                    \Sigma^{a,\mathrm{E}}_D(\tilde{\chi}^0_i)
   + \Sigma^{a,\mathrm{E}}_P(\tilde{\chi}^0_i)
       \Sigma^{a,\mathrm{O}}_D(\tilde{\chi}^0_i)\right],
\end{equation}
where the first (second) term is sensitive to
the CP phases in the production (decay)
of the neutralino $\tilde{\chi}^0_i$.
In the following we derive explicitly the T-odd
contributions to the spin density matrices of production and decay.

\subsection{T-odd terms of the production density matrix}

As can be seen from Eq.~(\ref{eq_kinsp4}) in Appendix B,
in $\Sigma^{a}_P(\tilde{\chi}^0_i)$ the expressions
$f_4^a$ vanish for $a=1,3\,$, 
because they do not contain three linearly independent
vectors, hence $\Sigma^{1,\mathrm{O}}_P(\tilde{\chi}^0_i) = 
\Sigma^{3,\mathrm{O}}_P(\tilde{\chi}^0_i) =0$.
Only for $a=2$ we get a T-odd contribution 
to the production density matrix 
$\Sigma^{2,\mathrm{O}}_P(\tilde{\chi}^0_i)$,
which is related to the polarisation
of the neutralino $\tilde{\chi}^0_i$
perpendicular to the production plane
(see the expression for $s^2(\tilde{\chi}^0_i)$, Eq.~(\ref{eq_22})
in Appendix A). We obtain from Eqs.~(\ref{eq_spzz}) -- (\ref{eq_spsubs})
for $a=2$
\begin{eqnarray}
\Sigma^{2,\mathrm{O}}_P(\tilde{\chi}^0_i) & = &
\Sigma^{2,\mathrm{O}}_P(Z Z) + \Sigma^{2,\mathrm{O}}_P(Z
\tilde{e}_L) +
\Sigma^{2,\mathrm{O}}_P(Z \tilde{e}_R) 
+\Sigma^{2,\mathrm{O}}_P(\tilde{e}_L \tilde{e}_L) +
\Sigma^{2,\mathrm{O}}_P(\tilde{e}_R \tilde{e}_R) \nonumber\\&&
\end{eqnarray}
with
\begin{eqnarray}
\Sigma_P^{2, \mathrm{O}}(ZZ)&=& \frac{g^4}{\cos^4\Theta_W} |\Delta^s(Z)|^2
    \Big( R^2_{\ell} (1+P_{e^-})(1-P_{e^+}) - 
      L^2_{\ell} (1-P_{e^-})(1+P_{e^+}) \Big) \nonumber\\
  & & \times \Big[ 2 (Re O^{''L}_{ij}) (Im O^{''L}_{ij})  i f_4^{2} \Big] ,
    \label{eq_spzz_to}\\
\Sigma_P^{2, \mathrm{O}}(Z\tilde{e}_L) 
  & = & \frac{g^4}{2 \cos^2\Theta_W} L_{\ell}
    (1-P_{e^-})(1+P_{e^+}) \nonumber\\
  & & \times Re\Big\{\Delta^s(Z) 
    \Big[f^{L*}_{\ell i}f^{L}_{\ell j} O^{''L*}_{ij}
    \Delta^{t*}(\tilde{e}_L) -f^{L}_{\ell i}f^{L*}_{\ell j}
    O^{''L}_{ij} \Delta^{u*}(\tilde{e}_L)
    \Big]f_4^{2}\Big\} , \label{eq_spzel_to}\\
\Sigma_P^{2, \mathrm{O}}(\tilde{e}_L \tilde{e}_L)
  & = & \frac{g^4}{4} (1-P_{e^-})(1+P_{e^+})
    Re\Big\{ \Delta^{u}(\tilde{e}_L)\Delta^{t*}(\tilde{e}_L)
    (f^{L*}_{\ell i})^2 (f^{L}_{\ell j})^2 f_4^{2}\Big\} ,
    \label{eq_spelel_to}\\
\Sigma_P^{2, \mathrm{O}}(Z\tilde{e}_R) 
  & = & \frac{g^4}{2 \cos^2\Theta_W} R_{\ell}
    (1+P_{e^-})(1-P_{e^+}) \nonumber\\
  & & \times Re\Big\{\Delta^s(Z) 
    \Big[f^{R*}_{\ell i}f^{R}_{\ell j} O^{''R*}_{ij}
    \Delta^{t*}(\tilde{e}_R) -f^{R}_{\ell i}f^{R*}_{\ell j}
    O^{''R}_{ij} \Delta^{u*}(\tilde{e}_R)
    \Big]f_4^{2}\Big\} , \label{eq_spzel_tor}\\
\Sigma_P^{2, \mathrm{O}}(\tilde{e}_R \tilde{e}_R)
  & = & \frac{g^4}{4} (1+P_{e^-})(1-P_{e^+})
    Re\Big\{ \Delta^{u}(\tilde{e}_R)\Delta^{t*}(\tilde{e}_R)
    (f^{R*}_{\ell i})^2 (f^{R}_{\ell j})^2 f_4^{2}\Big\} ,
    \label{eq_spelel_tor}
\end{eqnarray}
where $P_{e^-}$ and $P_{e^+}$ denotes the degree of 
longitudinal polarisation of 
the electron beam and positron beam, respectively.
The function $f^a_4$, Eq.~(\ref{eq_kinsp4}), for $a=2$ reads
\begin{equation}
f_4^{2}= i m_j \epsilon_{\mu\nu\rho\sigma} p_2^{\mu}
p_1^{\nu} s^{2 \rho} p_3^{\sigma}. \label{eq_f4_a=2}
\end{equation}
Note that $f^2_4$ is purely imaginary.
When inserted, for example, into the expression
for $\Sigma^{2,\mathrm{O}}_P(Z \tilde{e}_L)$,
Eq.~(\ref{eq_spzel_to}), $f^2_4$ is multiplied
by the factor $i\cdot Im\{f^L_{\ell i} f^{L*}_{\ell j} O^{''L}_{ij}\}$,
which is non-vanishing only if the couplings are complex.
Hence it gives a CP-sensitive contribution
to the asymmetry $A_T$, which depends on the phases
$\phi_\mu$ and $\phi_{M_1}$.
Analogous contributions come from the
other terms in $\Sigma^{2,\mathrm{O}}_P$, 
Eqs.~(\ref{eq_spzz_to}),(\ref{eq_spelel_to}) -- (\ref{eq_spelel_tor}).
We have to multiply $\Sigma^{2,\mathrm{O}}_P$ in
Eq.~(\ref{CPinT}) by $\Sigma^{2,\mathrm{E}}_D$,
for which we obtain from Eqs.~(\ref{eq_dssum}) -- (\ref{eq_dssub3}),
Appendix B,
\begin{equation}
\Sigma^{2, \mathrm{E}}_D(\tilde{\chi}^0_i) = 
\Sigma^{2,\mathrm{E}}_D(Z Z) + 
\Sigma^{2,\mathrm{E}}_D(Z \tilde{\ell}_L) +
\Sigma^{2,\mathrm{E}}_D(Z \tilde{\ell}_R) +
\Sigma^{2,\mathrm{E}}_D(\tilde{\ell}_L \tilde{\ell}_L) +
\Sigma^{2,\mathrm{E}}_D(\tilde{\ell}_R \tilde{\ell}_R)
\end{equation}
with
\begin{eqnarray}
\Sigma_D^{2, \mathrm{E}}(ZZ) & = & 
    8 \frac{g^4}{\cos^4\Theta_W} |\Delta^{s_i}(Z)|^2
    (R^2_{\ell}-L_{\ell}^2) \nonumber\\
  & & \times\Big[ -[(Re O^{''L}_{ki})^2 -(Im O^{''L}_{ki})^2]g^{2}_3+
    |O^{''L}_{ki}|^2(g^{2}_1-g^{2}_2)\Big],
    \label{eq_dszz_te}\\
\Sigma_D^{2, \mathrm{E}}(Z \tilde{\ell}_L) & = & \frac{4
    g^4}{\cos^2\Theta_W}L_{\ell}
    Re\Big\{\Delta^{s_i}(Z)
    \Big[ f^L_{\ell i} f^{L*}_{\ell k} \Delta^{t_i*}(\tilde{\ell}_L)
    \big(-2 O^{''L}_{ki} g^{2}_1 +O^{''L*}_{ki} g^{2}_3\big)\nonumber\\
  & & \phantom{\frac{4 g^4}{\cos^2\Theta_W}L_{\ell} Re\Big\{\Delta^{s_i}(Z)}
    + f^{L*}_{\ell i} f^{L}_{\ell k} \Delta^{u_i*}(\tilde{\ell}_L)
    \big(2 O^{''L*}_{ki} g^{2}_2 +O^{''L}_{ki} g^{2}_3\big)
    \Big]\Big\},\label{eq_dszel_te}\\
\Sigma_D^{2, \mathrm{E}}(\tilde{\ell}_L \tilde{\ell}_L)&=& 2 g^4 \Big[
    |f^{L}_{\ell i}|^2 |f^L_{\ell k}|^2
    [|\Delta^{u_i}(\tilde{\ell}_L)|^2 g_2^{2}
    -|\Delta^{t_i}(\tilde{\ell}_L)|^2 g_1^{2}]\nonumber\\
  & & \phantom{2 g^4 \Big[}
    + Re\big\{ (f^{L*}_{\ell i})^2 (f^L_{\ell k})^2
    \Delta^{t_i}(\tilde{\ell}_L)
    \Delta^{u_i*}(\tilde{\ell}_L)g_3^{2}\big\}\Big],\label{eq_dselel_te}\\
\Sigma_D^{2, \mathrm{E}}(Z \tilde{\ell}_R)&=& \frac{4
    g^4}{\cos^2\Theta_W}R_{\ell}
    Re\Big\{\Delta^{s_i}(Z)
    \Big[ f^R_{\ell i} f^{R*}_{\ell k} \Delta^{t_i*}(\tilde{\ell}_R)
    \big(2 O^{''R}_{ki} g^{2}_1 -O^{''R*}_{ki} g^{2}_3\big)\nonumber\\
  & & \phantom{\frac{4 g^4}{\cos^2\Theta_W}R_{\ell}\Delta^{s_i}(Z)}
    + f^{R*}_{\ell i} f^{R}_{\ell k} \Delta^{u_i*}(\tilde{\ell}_R)
    \big(-2 O^{''R*}_{ki} g^{2}_2 -O^{''R}_{ki} g^{2}_3\big)
    \Big]\Big\},\label{eq_dszel_ter}\\
\Sigma_D^{2, \mathrm{E}}(\tilde{\ell}_R \tilde{\ell}_R)&=& 2 g^4 \Big[
    |f^{R}_{\ell i}|^2 |f^R_{\ell k}|^2
    [-|\Delta^{u_i}(\tilde{\ell}_R)|^2 g_2^{2}
    +|\Delta^{t_i}(\tilde{\ell}_R)|^2 g_1^{2}]\nonumber\\
  & & \phantom{2 g^4 \Big[}
    - Re\big\{ (f^{R*}_{\ell i})^2 (f^R_{\ell k})^2
    \Delta^{t_i}(\tilde{\ell}_R)
    \Delta^{u_i*}(\tilde{\ell}_R)g_3^{2}\big\}\Big],\label{eq_dselel_ter}
\end{eqnarray}
where
\begin{eqnarray}
g^{2}_1&=& m_i (p_5 p_7) (p_6 s^2), \label{eq_g1_a=2}\\
g^{2}_2&=& m_i (p_5 p_6) (p_7 s^2), \label{eq_g2+a=2}\\
g^{2}_3&=& m_k [(p_3 p_6) (p_7 s^2)-(p_3 p_7) (p_6 s^2)].
\label{eq_g3_a=2}
\end{eqnarray}
The kinematic functions $g^2_1$, $g^2_2$, $g^2_3$ are real.
When multiplied by the purely imaginary $f^2_4$, Eq.~(\ref{eq_f4_a=2}),
this leads to triple products sensitive to the CP phases in
the production process,
which in the laboratory system read: 
\begin{eqnarray}
g^{2}_1\cdot f_4^{2} &=& i 2 E_b m_i m_j (p_5 p_7) 
\vec{p}_6 ( \vec{p}_1\times \vec{p}_3) , \label{eq_coupl_1}\\
g^{2}_2\cdot f_4^{2} &=& i 2 E_b m_i m_j (p_5 p_6) 
\vec{p}_7 ( \vec{p}_1\times \vec{p}_3) , \label{eq_coupl_2}\\
g^{2}_3\cdot f_4^{2} &=& i 2 E_b m_j m_k \{ 
 (p_3 p_6)\vec{p}_7 (\vec{p}_1 \times \vec{p}_3)
-(p_3 p_7)\vec{p}_6 (\vec{p}_1 \times \vec{p}_3) \}.
\label{eq_coupl_3}
\end{eqnarray}
As outlined above, these expressions will be 
multiplied in Eqs.~(\ref{eq_spzz_to}) -- (\ref{eq_spelel_tor})
by the factors
$i\cdot Im\{f^L_{\ell i} f^{L*}_{\ell j} O^{''L}_{ij}\}$
etc. and contribute to the first term of 
Eq.~(\ref{CPinT}) and, hence,
to the numerator of the asymmetry
$A_T$, Eq.~(\ref{Asy}).

\subsection{T-odd terms of the decay density matrix}
The second term in Eq.~(\ref{CPinT}) is sensitive
to CP violation in the neutralino decay process (\ref{process2}).
The expressions $g_4^a$, Eq.~(\ref{eq_dssub6}),
in $\Sigma^{a,\mathrm{O}}_D(\tilde{\chi}^0_i)$
contain three linearly independent vectors, however,
$\Sigma^{2,\mathrm{E}}_P(\tilde{\chi}^0_i)$ (Eq.~(\ref{eq_sumdecay_te}))
vanishes because 
$s^2(\tilde{\chi}^0_i)$ is perpendicular to the scattering plane.
Therefore, only the terms
$\Sigma^{a,\mathrm{E}}_P(\tilde{\chi}^0_i)$ for
$a=1,3$ have to be taken into account.
The CP-sensitive terms of the decay density matrix following
from Eqs.~(\ref{eq_dssum}) -- (\ref{eq_dssub3}) for $a=1,3$, are
\begin{equation}
\Sigma^{a,\mathrm{O}}_D(\tilde{\chi}^0_i) =
\Sigma^{a,\mathrm{O}}_D(Z Z) +
\Sigma^{a,\mathrm{O}}_D(Z \tilde{\ell}_L) +
\Sigma^{a,\mathrm{O}}_D(Z \tilde{\ell}_R) +
\Sigma^{a,\mathrm{O}}_D(\tilde{\ell}_L \tilde{\ell}_L) +
\Sigma^{a,\mathrm{O}}_D(\tilde{\ell}_R \tilde{\ell}_R)
\label{eq_todecay}
\end{equation}
with
\begin{eqnarray}
\Sigma_D^{a, \mathrm{O}}(ZZ) &=& 
    8 \frac{g^4}{\cos^4\Theta_W} |\Delta^{s_i}(Z)|^2
    (L^2_{\ell} - R_{\ell}^2) 
    \Big[ 2 Re(O^{''L}_{ki}) Im(O^{''L}_{ki}) i g_4^a \Big],
    \label{eq_dszz_to}\\
\Sigma_D^{a, \mathrm{O}}(Z \tilde{\ell}_L) & = & 
    \frac{4 g^4}{\cos^2\Theta_W}L_{\ell}
    Re\Big\{\Delta^{s_i}(Z)
    \Big[-f^L_{\ell i} f^{L*}_{\ell k} O^{''L*}_{ki}
    \Delta^{t_i*}(\tilde{\ell}_L)\nonumber\\
  & &\phantom{\frac{4 g^4}{\cos^2\Theta_W}L_{\ell}Re\Big\{\Delta^{s_i}(Z)\Big[}
    + f^{L*}_{\ell i} f^{L}_{\ell k} O^{''L}_{ki}
    \Delta^{u_i*}(\tilde{\ell}_L)
    \Big]g_4^a\Big\},\label{eq_dszel_to}\\
\Sigma_D^{a, \mathrm{O}}(\tilde{\ell}_L \tilde{\ell}_L)&=& 
    2 g^4 Re\Big\{ (f^{L*}_{\ell i})^2 (f^L_{\ell k})^2
    \Delta^{t_i}(\tilde{\ell}_L)
    \Delta^{u_i*}(\tilde{\ell}_L) g_4^a\Big\},\label{eq_dselel_to}\\
\Sigma_D^{a, \mathrm{O}}(Z \tilde{\ell}_R) & = & 
    \frac{4 g^4}{\cos^2\Theta_W} R_{\ell}
    Re\Big\{\Delta^{s_i}(Z)
    \Big[-f^R_{\ell i} f^{R*}_{\ell k} O^{''R*}_{ki}
    \Delta^{t_i*}(\tilde{\ell}_R)\nonumber\\
  & &\phantom{\frac{4 g^4}{\cos^2\Theta_W}R_{\ell}Re\Big\{\Delta^{s_i}(Z)\Big[}
    + f^{R*}_{\ell i} f^{R}_{\ell k} O^{''L}_{ki}
    \Delta^{u_i*}(\tilde{\ell}_R)
    \Big]g_4^a\Big\},\label{eq_dszel_tor}\\
\Sigma_D^{a, \mathrm{O}}(\tilde{\ell}_R \tilde{\ell}_R) & = & 
    2 g^4 Re\Big\{ (f^{R*}_{\ell i})^2 (f^R_{\ell k})^2
    \Delta^{t_i}(\tilde{\ell}_R)
    \Delta^{u_i*}(\tilde{\ell}_R) g_4^a\Big\},\label{eq_dselel_tor}
\end{eqnarray}
where for $a=1,3$ we have
\begin{equation}
g_4^a=i m_k \epsilon_{\mu\nu\rho\sigma} s^{a\mu} p_3^{\nu}
p_7^{\rho} p_6^{\sigma}. \label{g4-decay}
\end{equation}
Now $g^a_4$, $a=1,3\,$, is purely imaginary.
When inserted, for example, in Eq.~(\ref{eq_dszel_to})
it is multiplied by the factor
$i\cdot Im\{f^L_{\ell i} f^{L*}_{\ell k} O^{''L*}_{ki}\}$,
which depends on the phases $\phi_\mu$ and $\phi_{M_1}$
and contributes to $\Sigma_D^{a, \mathrm{O}}$.
Analogous contributions follow from 
Eqs.~(\ref{eq_dszz_to}),(\ref{eq_dselel_to}) -- (\ref{eq_dselel_tor}).
The corresponding T-even terms of the production
density matrix also entering in Eq.~(\ref{CPinT}) are
obtained from Eqs.~(\ref{eq_spzz}) -- (\ref{eq_spsubs}),
\begin{equation}
\Sigma_P^{a, \mathrm{E}}(\tilde{\chi}^0_i)=
 \Sigma_P^{a, \mathrm{E}}(ZZ)
+\Sigma_P^{a, \mathrm{E}}(Z\tilde{e}_L) +\Sigma_P^{a, \mathrm{E}}
(Z\tilde{e}_R)
+\Sigma_P^{a, \mathrm{E}}(\tilde{e}_L \tilde{e}_L) +
\Sigma_P^{a, \mathrm{E}}(\tilde{e}_R\tilde{e}_R),\label{eq_sumdecay_te}
\end{equation}
where
\begin{eqnarray}
\Sigma_P^{a, \mathrm{E}}(ZZ)&=& \frac{g^4}{\cos^4\Theta_W} |\Delta^s(Z)|^2
    \Big( R^2_{\ell} (1+P_{e^-})(1-P_{e^+})
      - L^2_{\ell} (1-P_{e^-})(1+P_{e^+}) \Big) \nonumber\\
  & & \times \Big[ |O^{''L}_{ij}|^2 (f_2^a-f_1^a)
    - [(Re O^{''L}_{ij})^2 -(Im O^{''L}_{ij})^2] f_3^a \Big] ,
    \label{eq_spzz_te}\\
\Sigma_P^{a, \mathrm{E}}(Z\tilde{e}_L) 
  & = & \frac{g^4}{2 \cos^2\Theta_W} L_{\ell}
    (1-P_{e^-})(1+P_{e^+}) \nonumber\\
  & & \times Re\Big\{\Delta^s(Z) \Big[2 f^{L}_{\ell i}f^{L*}_{\ell
    j} O^{''L*}_{ij} \Delta^{u*}(\tilde{e}_L)f_1^a -2 f^{L*}_{\ell
    i}f^{L}_{\ell j} O^{''L}_{ij}
    \Delta^{t*}(\tilde{e}_L)f_2^a \nonumber\\
  & & \phantom{\times Re\Big\{\Delta^s(Z)} 
    +\big[f^{L}_{\ell i}f^{L*}_{\ell j} O^{''L}_{ij}
    \Delta^{u*}(\tilde{e}_L) +f^{L*}_{\ell i}f^{L}_{\ell j}
    O^{''L*}_{ij} \Delta^{t*}(\tilde{e}_L)\big]
    f_3^a\Big]\Big\} , \label{eq_spzel_te}\\
\Sigma_P^{a, \mathrm{E}}(\tilde{e}_L \tilde{e}_L)
  & = & \frac{g^4}{4} (1-P_{e^-})(1+P_{e^+})
    \Big[
    |f^L_{\ell i}|^2 |f^L_{\ell j}|^2 \big( |\Delta^u
    (\tilde{e}_L)|^2 f_1^a-|\Delta^t (\tilde{e}_L)|^2 f_2^a
    \big) \nonumber\\
  & & \phantom{\frac{g^4}{4} (1-P_{e^-})(1+P_{e^+})}
    + Re\{ (f^{L*}_{\ell i})^2 (f^{L}_{\ell j})^2
    \Delta^{u}(\tilde{e}_L)\Delta^{t*}(\tilde{e}_L) f_3^a\}\Big] ,
    \label{eq_spelel_te}\\
\Sigma_P^{a, \mathrm{E}}(Z\tilde{e}_R) 
  & = & \frac{g^4}{2 \cos^2\Theta_W} R_{\ell}
    (1+P_{e^-})(1-P_{e^+})
    \nonumber\\
  & & \times Re\Big\{\Delta^s(Z) \Big[-2 f^{R}_{\ell i}f^{R*}_{\ell
    j} O^{''R*}_{ij} \Delta^{u*}(\tilde{e}_R)f_1^a +2 f^{R*}_{\ell
    i}f^{R}_{\ell j} O^{''R}_{ij}
    \Delta^{t*}(\tilde{e}_R)f_2^a \nonumber\\
  & & \phantom{Re\Big\{\Delta^s(Z)}
    - \big[f^{R}_{\ell i}f^{R*}_{\ell j} O^{''R}_{ij}
    \Delta^{u*}(\tilde{e}_R) +f^{R*}_{\ell i}f^{R}_{\ell j}
    O^{''R*}_{ij} \Delta^{t*}(\tilde{e}_R)\big]
    f_3^a\Big]\Big\} , \label{eq_spzel_ter}\\
\Sigma_P^{a, \mathrm{E}}(\tilde{e}_R \tilde{e}_R)
  & = & \frac{g^4}{4} (1+P_{e^-})(1-P_{e^+})
    \Big[
    |f^R_{\ell i}|^2 |f^R_{\ell j}|^2 \big(-|\Delta^u
    (\tilde{e}_R)|^2 f_1^a+|\Delta^t (\tilde{e}_R)|^2 f_2^a
    \big) \nonumber\\
  & & \phantom{\frac{g^4}{4} (1+P_{e^-})(1-P_{e^+})}
    - Re\{ (f^{R*}_{\ell i})^2 (f^{R}_{\ell j})^2
    \Delta^{u}(\tilde{e}_R)\Delta^{t*}(\tilde{e}_R) f_3^a\}\Big],
    \label{eq_spelel_ter}
\end{eqnarray}
where
\begin{eqnarray}
f^a_1&=& m_i (p_2 p_4)(p_1 s^a),\label{eq_kinsp1-te}\\
f^a_2&=& m_i (p_1 p_4)(p_2 s^a),\label{eq_kinsp2-te}\\
f^a_3&=& m_j [(p_1 p_3)(p_2 s^a)-(p_2 p_3)(p_1 s^a)].
\label{eq_kinsp3-te}
\end{eqnarray}
The triple products sensitive to the CP phases
in the decay read in the laboratory system:
\begin{eqnarray}
\sum_{a=1,3} f^a_1 \cdot g^a_4 &=&
  i m_i m_k (p_2 p_4)\Big\{-E_b \vec{p}_5 (\vec{p}_7\times \vec{p}_6)
  -E_7 \vec{p}_5 (\vec{p}_6\times \vec{p}_1 ) \nonumber\\[-3mm]
 &&\phantom{i m_i m_k (p_2 p_4)\Big\{}
  + E_6 \vec{p}_5 (\vec{p}_7\times \vec{p}_1)
  + E_5 \vec{p}_1 (\vec{p}_7\times \vec{p}_6) \Big\} , \label{eq_te1}\\[2mm]
\sum_{a=1,3} f^a_2 \cdot g^a_4 &=&
  i m_i m_k (p_1 p_4)\Big\{-E_b \vec{p}_5 (\vec{p}_7\times \vec{p}_6)
  +E_7 \vec{p}_5 (\vec{p}_6\times \vec{p}_1 ) \nonumber\\[-3mm]
 &&\phantom{i m_i m_k (p_2 p_4)\Big\{}
  - E_6 \vec{p}_5 (\vec{p}_7\times \vec{p}_1)
  - E_5 \vec{p}_1 (\vec{p}_7\times \vec{p}_6) \Big\} , \label{eq_te2}\\[2mm]
\sum_{a=1,3} f^a_3 \cdot g^a_4 &=&
  i m_j m_k \Big\{[(p_2 p_3)-(p_1 p_3)]
  E_b \vec{p}_5 (\vec{p}_7\times \vec{p}_6)\nonumber\\[-3mm]
 &&\phantom{i m_j m_k \Big\{} + [(p_2 p_3)+(p_1 p_3)]\nonumber\\
 &&\phantom{i m_j m_k \Big\{ +\;} \Big[E_7 \vec{p}_5 
  (\vec{p}_6\times \vec{p}_1 )
  -E_6 \vec{p}_5 (\vec{p}_7\times \vec{p}_1)
  -E_5 \vec{p}_1 (\vec{p}_7\times \vec{p}_6) \Big] \Big\} .
  \label{eq_te3}
\end{eqnarray}
As already mentioned, these quantities will be 
multiplied in Eqs.~(\ref{eq_dszz_to}) -- (\ref{eq_dselel_tor})
by the factor
$i\cdot Im\{f^L_{\ell i} f^{L*}_{\ell k} O^{''L*}_{ki}\}$
etc. and contribute to the second term of
Eq.~(\ref{CPinT}).

We take into account only the contributions to the T-odd asymmetry
$A_T$ which stem from the complex couplings and neglect the
contributions from the widths of $Z$, $\tilde{\ell}_L$ and 
$\tilde{\ell}_R$ in the
propagators Eqs.~(\ref{eq_Zprop}), (\ref{eq_selprop}),
because they are of higher order.

\section{Numerical Results}
In this section
we analyse numerically the CP asymmetry $A_T$, Eq.~(\ref{Asy}),
and the cross section for reactions (\ref{process1}) and 
(\ref{process2}),
at an $e^+e^-$ linear collider with $\sqrt{s} = 500$~GeV and
longitudinally polarised $e^\pm$ beams.
In particular, we investigate the dependence on 
the phases of the complex parameters
$M_1=|M_1|e^{i\phi_{M_1}}$ and $\mu=|\mu|e^{i\phi_{\mu}}$.
For our numerical analysis we choose two scenarios A and B, such that 
$m_{\tilde{\chi}^0_2} < m_{\tilde{\chi}^0_1}+m_{Z^0}$
and $m_{\tilde{\chi}^0_2}< m_{\tilde{\ell}_{L,R}}$
to prevent two-body decays of $\tilde{\chi}^0_2$.
The SUSY parameters and the 
masses of $\tilde{\chi}^0_i$, $i=1,\ldots,4$,
and $\tilde{\ell}_L$, $\tilde{\ell}_R$ are given in Table 1.
The decay widths of the neutralinos have been computed with help of
the program SPheno \cite{Porod:2003um}.

\begin{table}[H]
\renewcommand{\arraystretch}{1.3}
\begin{center}
\begin{tabular}{|c||c|c||c|c|} \hline
 Scenario & \multicolumn{2}{c||}{A} & \multicolumn{2}{c|}{B}\\
\hline\hline
 $|M_1|$ & \multicolumn{2}{c||}{150} & \multicolumn{2}{c|}{90}\\ \hline
 $M_2$ & \multicolumn{2}{c||}{300} & \multicolumn{2}{c|}{192.7}\\ \hline
 $|\mu|$ & \multicolumn{2}{c||}{200} & \multicolumn{2}{c|}{352.4}\\ \hline
 $\tan{\beta}$ & \multicolumn{2}{c||}{10} & \multicolumn{2}{c|}{10}\\
\hline
 $m_{\tilde{\ell}_L}$ & \multicolumn{2}{c||}{267.6} &
   \multicolumn{2}{c|}{267.6}\\ \hline
 $m_{\tilde{\ell}_R}$ & \multicolumn{2}{c||}{224.4} &
   \multicolumn{2}{c|}{224.4}\\
\hline
 $(\phi_{M_1}, \phi_{\mu})$ & 
   \makebox[20mm]{$(0.5\pi,0)$} & \makebox[20mm]{$(0.5\pi,0.5\pi)$} &
   \makebox[20mm]{$(0.5\pi,0)$} & \makebox[20mm]{$(0.5\pi,0.5\pi)$} \\ \hline
 $m_{\tilde{\chi}^0_1}$ & 135.4 & 139.9 & 88.5 & 89.7 \\ \hline
 $m_{\tilde{\chi}^0_2}$ & 182.2 & 182.8 & 176.2 & 180.3 \\ \hline
 $m_{\tilde{\chi}^0_3}$ & 213.3 & 214.9 & 359.4 & 360.7 \\ \hline
 $m_{\tilde{\chi}^0_4}$ & 334.8 & 331.6 & 377.1 & 373.6 \\ \hline
\end{tabular}\\[0.5ex]
\caption{\label{scentab}
Input parameters $|M_1|$, $M_2$, $|\mu|$, $\tan\beta$,
$m_{\tilde{\ell}_L}$ and $m_{\tilde{\ell}_R}$
and the resulting masses $m_{\tilde{\chi}^0_i}$, $i=1,\ldots,4$,
for $(\phi_{M_1},\phi_{\mu}) = (0.5\pi,0)$ and $(0.5\pi,0.5\pi)$.
The parameters $M_2$, $|\mu|$ and $\tan\beta$ in scenario B
are chosen according to the scenario SPS1a in \cite{sps}.
All masses are given in GeV.}
\end{center}
\end{table}

\begin{figure}[ht!]
\begin{picture}(16,5.9)
\put(0.1,5.5){$\Gamma(\tilde{\chi}^0_2)/$keV}
\put(8.4,5.5){$BR(\tilde{\chi}^0_2 \to \tilde{\chi}^0_1 \ell^+ \ell^-)$
  in \%}
\put(14.85,0.1){$\phi_{M_1}/\pi$}
\put(6.7,0.1){$\phi_{M_1}/\pi$}
\put(0.8,4.8){(a)}
\put(9.3,4.8){(b)}
\put(0,0.4){\epsfig{file=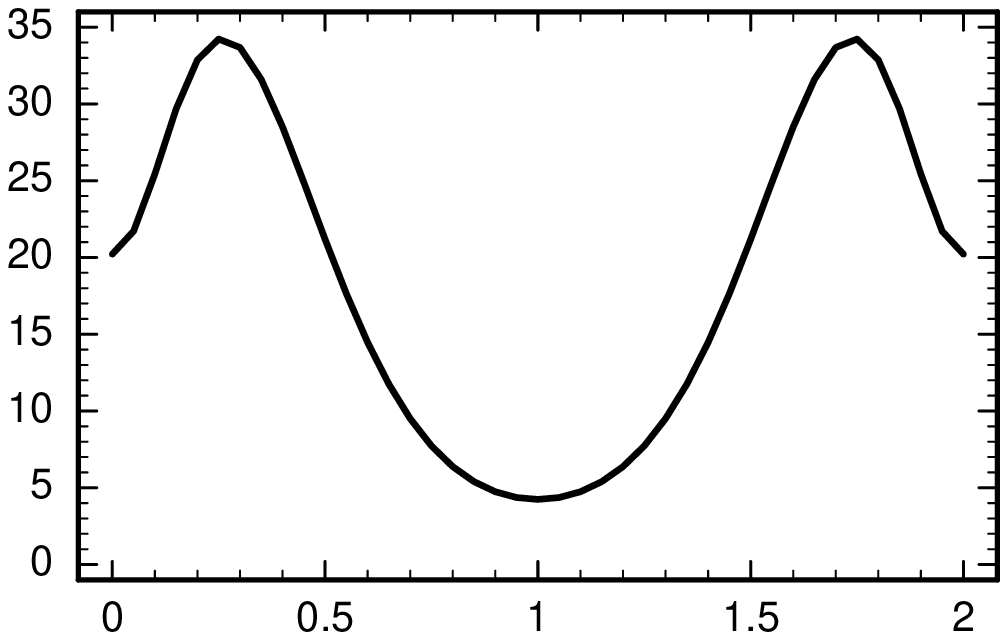,scale=0.76}}
\put(8.35,0.4){\epsfig{file=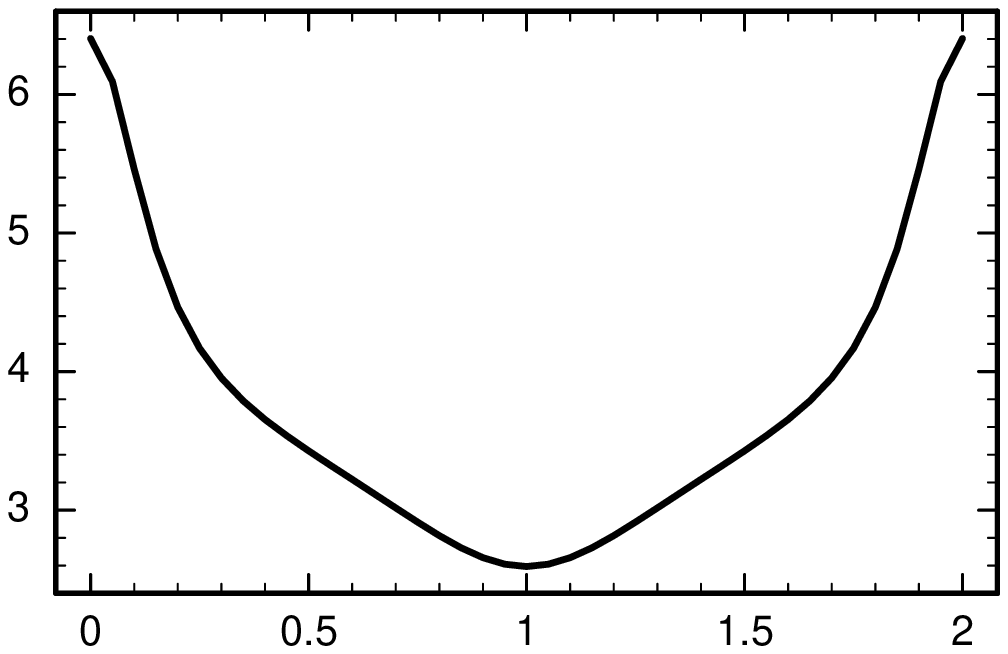,scale=0.745}}
\end{picture}
\caption{\label{fig:gamchi2}
(a) Total decay width $\Gamma(\tilde{\chi}^0_2)$ and (b) branching ratio
$BR(\tilde{\chi}^0_2 \to \tilde{\chi}^0_1 \ell^+ \ell^-)$, $\ell=e$ or
$\mu$, in scenario A of Table~\ref{scentab} for $\phi_\mu=0$.}
\end{figure}

In scenario A the neutralinos
$\tilde{\chi}^0_1$, $\tilde{\chi}^0_2$, $\tilde{\chi}^0_3$
are mainly $\tilde{B}$-higgsino, higgsino-$\tilde{B}$-$\tilde{W}^3$, 
higgsino-$\tilde{B}$ mixtures, respectively, where
the $\tilde{B}$ component of the $\tilde{\chi}^0_2$ has a
relatively strong $\phi_{M_1}$ dependence.
In Fig.~\ref{fig:gamchi2} we show the total decay width
$\Gamma(\tilde{\chi}^0_2)$ and the branching ratio 
$BR(\tilde{\chi}^0_2 \to \tilde{\chi}^0_1 \ell^+ \ell^-)$, $\ell = e$ or
$\mu$ in scenario A as a function of $\phi_{M_1}$ for $\phi_\mu=0$.
As can be seen, the CP-even quantities
$\Gamma(\tilde{\chi}^0_2)$ and 
$BR(\tilde{\chi}^0_2 \to \tilde{\chi}^0_1 \ell^+ \ell^-)$ 
depend quite strongly on $\phi_{M_1}$.
$\Gamma(\tilde{\chi}^0_2)$ varies between 34~keV for
$\phi_{M_1}=0.25\pi$, $1.75\pi$ and 4.2~keV for $\phi_{M_1}=\pi$, and
$BR(\tilde{\chi}^0_2 \to \tilde{\chi}^0_1 \ell^+ \ell^-)$ 
between 6.4\,\% for $\phi_{M_1}=0, 2\pi$ and 2.6\,\% for $\phi_{M_1}=\pi$.
The parameters $M_2$, $|\mu|$ and $\tan\beta$
in scenario B are chosen like in the scenario 
SPS1a in \cite{sps}, with 
$\tilde{\chi}^0_1$, $\tilde{\chi}^0_2$, $\tilde{\chi}^0_3$
being mainly $\tilde{B}$-, $\tilde{W}^3$-, higgsino-like,
respectively.
In this scenario $\Gamma(\tilde{\chi}^0_2)$ varies between
80~keV for $\phi_{M_1}=0$, $2\pi$ and 9.9~keV for $\phi_{M_1}=\pi$,
and $B(\tilde{\chi}^0_2 \to \tilde{\chi}^0_1 \ell^+ \ell^-)$ between
1.9\,\% for $\phi_{M_1}=0.4\pi$, $1.6\pi$ and 2.9\,\% for $\phi_{M_1}=\pi$.
In both scenarios we take for the slepton masses
$m_{\tilde{\ell}_L}=267.6$~GeV,
$m_{\tilde{\ell}_R}=224.4$~GeV and 
for the squark masses $m_{\tilde{u}_R}=597.6$~GeV,
$m_{\tilde{u}_L}=599.0$~GeV and
$m_{\tilde{d}_R}=600.5$~GeV, $m_{\tilde{d}_L}=602.9$~GeV 
for the first and second generation,
and $m_{\tilde{b}_1}=587.3$~GeV, $m_{\tilde{b}_2}=615.7$~GeV
($A_b=1000$~GeV) for the bottom squarks.
In most of the examples we take values
for the phases $\phi_\mu$ and $\phi_{M_1}$ which are
in agreement with the constraints from the electron
and neutron EDMs. In order to show the full phase dependence
of the asymmetry and the cross section, we also give
some examples where we vary $\phi_\mu$ and $\phi_{M_1}$
in the whole range, relaxing the constraints from the EDMs.  

\subsection{\boldmath $e^+e^- \to \tilde{\chi}^0_1\tilde{\chi}^0_2$}

\begin{figure}[ht!]
\begin{picture}(16,5.8)
\put(0.3,5.4){$A_T$ in \%}
\put(8.5,5.4){$\sigma/$fb}
\put(14.85,0.1){$\phi_{M_1}/\pi$}
\put(6.7,0.1){$\phi_{M_1}/\pi$}
\put(3.5,4.7){(a)}
\put(9.3,4.7){(b)}
\put(0,0.4){\epsfig{file=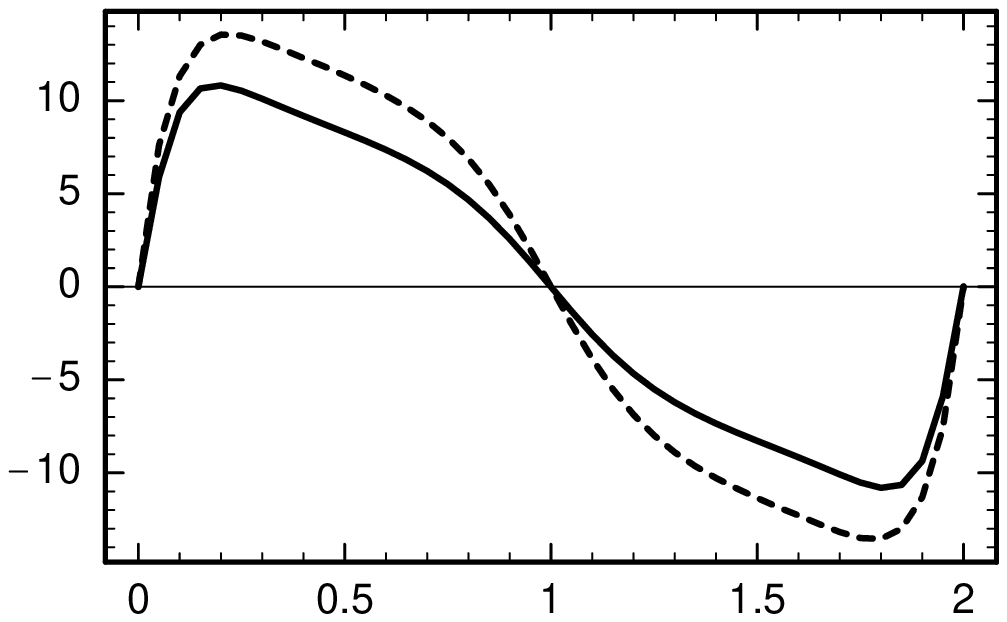,scale=0.77}}
\put(8.35,0.4){\epsfig{file=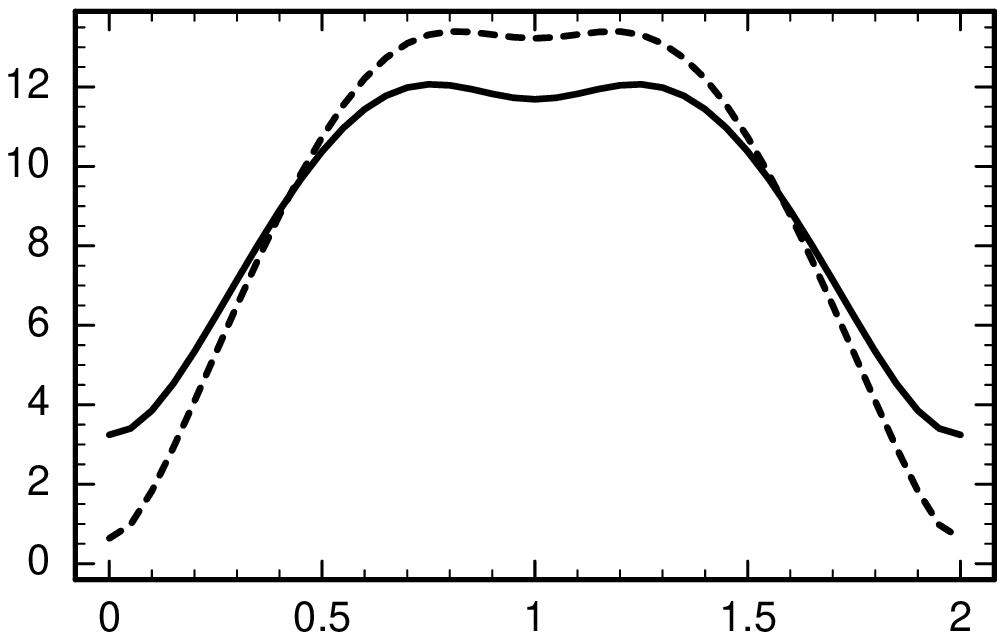,scale=0.75}}
\end{picture}
\caption{\label{fig:At12}
(a) CP asymmetry $A_T$ (Eq.~(\ref{Asy})) and (b) cross section
$\sigma(e^+e^- \to \tilde{\chi}^0_1\tilde{\chi}^0_2)\cdot
BR(\tilde{\chi}^0_2\rightarrow\tilde{\chi}^0_1 \ell^+\ell^-)$,
summed over $\ell=e,\mu\,$, 
in scenario A of Table~\ref{scentab} for $\phi_{\mu}=0$,
$P_{e^-}=-0.8$, $P_{e^+}=+0.6$ and
$\sqrt{s}=500$ GeV (solid), $\sqrt{s}=350$ GeV (dashed).}
\end{figure}

First we discuss neutralino production
$e^+e^-\to\tilde{\chi}^0_1\tilde{\chi}^0_2$ with
subsequent leptonic three-body decay
$\tilde{\chi}^0_2\to \tilde{\chi}^0_1\ell^+\ell^-$.
In this case one obtains CP-violating contributions from both production
and decay.
In Fig.~\ref{fig:At12} (a) we show the CP asymmetry $A_T$
as a function of the phase $\phi_{M_1}$
for scenario A defined in Table~\ref{scentab} for
$\phi_\mu = 0$ and two centre of mass energies
$\sqrt{s} = 350$~GeV and $\sqrt{s} = 500$~GeV.
The beam polarisations
are fixed at $P_{e^-}=-0.8$ and $P_{e^+}=+0.6$.
The CP asymmetry attains the largest values
of about 11\,\% ($-11\,\%$)
for $\sqrt{s} = 500$~GeV at $\phi_{M_1}\approx 0.2\pi$
($\phi_{M_1}\approx 1.8\pi$).
For $\sqrt{s} = 350$~GeV, i.e.\ closer to threshold of
the production,
$A_T$ reaches values of
$\pm 13.5\,\%$ at $\phi_{M_1}\approx 0.2\pi, 1.8\pi$.
Choosing $\phi_{\mu}=0.1\pi$ instead of $\phi_{\mu}=0$
leaves the asymmetry $A_T$ almost unchanged.
Note that $A_T$ does not have its 
largest values
at $\phi_{M_1}=\frac{\pi}{2}$ ($\phi_{M_1}=\frac{3\pi}{2}$),
where the imaginary parts of the respective couplings are maximal,
but at smaller (larger) phases.
The reason is an interplay between the $\phi_{M_1}$ dependence of the
cross section, shown in Fig.~\ref{fig:At12} (b),
and the $\phi_{M_1}$ dependence of 
the numerator of $A_T$, Eq.~(\ref{CPinT}).
This numerator is a sum of products of
CP-odd and CP-even factors, which essentially 
have a $\sin\phi_{M_1}$ and $\cos\phi_{M_1}$
behaviour, respectively (for $\phi_\mu=0$).
We plot in Fig.~\ref{fig:At12} (b) the cross section
$\sigma=\sigma(e^+e^- \to \tilde{\chi}^0_1\tilde{\chi}^0_2)\cdot 
BR(\tilde{\chi}^0_2\rightarrow\tilde{\chi}^0_1 \ell^+\ell^-)$
for the production and subsequent
decay process $e^+e^- \to \tilde{\chi}^0_1\tilde{\chi}^0_2
\to \tilde{\chi}^0_1\tilde{\chi}^0_1 \ell^+\ell^-$
(Eq.~(\ref{crosss})), summed over $\ell=e,\mu$. 
Note that also the cross section 
has a rather strong $\phi_{M_1}$ dependence.
At $\phi_{M_1}=\pi$ the production cross section 
$\sigma(e^+e^- \to \tilde{\chi}^0_1\tilde{\chi}^0_2)$
has a maximum, whereas the branching ratio
$BR(\tilde{\chi}^0_2\rightarrow\tilde{\chi}^0_1 \ell^+\ell^-)$
has a minimum (see Fig.~\ref{fig:gamchi2} (b)) resulting in the dip of 
the cross section for the combined process of production and decay
at $\phi_{M_1}=\pi$.

\begin{figure}[t]
\begin{picture}(16,8.6)

\put(0.2,8.3){$\phi_{M_1}/\pi$}
\put(8.4,8.3){$\phi_{M_1}/\pi$}
\put(6.9,0.1){$\phi_{\mu}/\pi$}
\put(15.1,0.1){$\phi_{\mu}/\pi$}

\put(1,6.2){(a)}
\put(9.2,6.9){(b)}

\put(4.1,7.75){\tiny 0}
\put(4.15,7.41){\tiny $-5$}
\put(4.05,7.1){\tiny $-10$}
\put(4.15,5.95){\tiny $-5$}
\put(4.3,4.3){\tiny 0}
\put(4.3,2.85){\tiny 5}
\put(4.2,1.7){\tiny 10}
\put(4.3,1.33){\tiny 5}
\put(4.3,1.1){\tiny 0}

\put(0,0.4){\epsfig{file=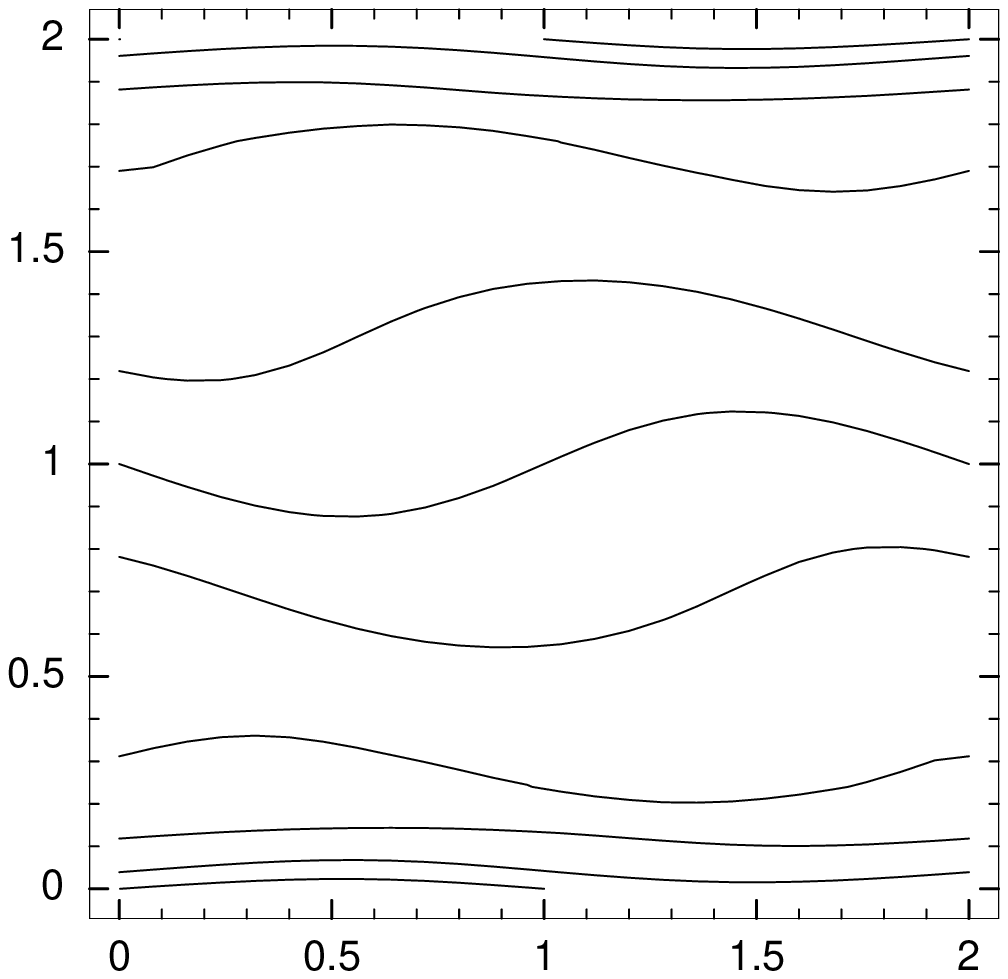,scale=0.77}}

\put(12.3,7.75){\tiny 0}
\put(12.5,7.3){\tiny 5}
\put(13.4,6.95){\tiny 7.5}
\put(12.5,6.45){\tiny 5}
\put(12.5,4.3){\tiny 0}
\put(12.3,2.35){\tiny $-5$}
\put(11,1.85){\tiny $-7.5$}
\put(12.3,1.55){\tiny $-5$}
\put(12.5,1.1){\tiny 0}

\put(8.2,0.4){\epsfig{file=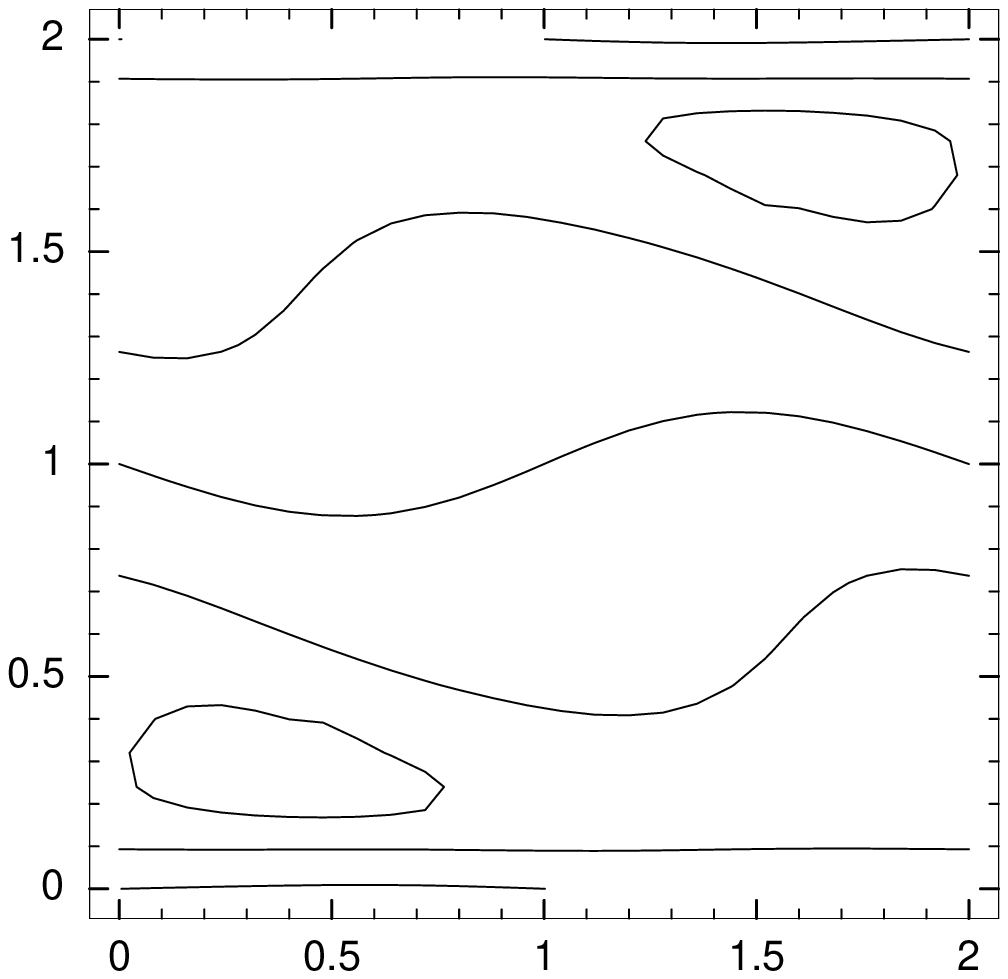,scale=0.77}}

\end{picture}
\caption{\label{fig:At12phiM1phimu}
Contours of the CP asymmetry $A_T$ (Eq.~(\ref{Asy})) in \%
for $e^+e^- \to \tilde{\chi}^0_1\tilde{\chi}^0_2$
with subsequent decay
$\tilde{\chi}^0_2 \to \tilde{\chi}^0_1\ell^+\ell^-$
in scenario A of Table~\ref{scentab} for
$\sqrt{s}=500$~GeV and (a) $P_{e^-}=-0.8$, $P_{e^+}=+0.6$ and
(b) $P_{e^-}=+0.8$, $P_{e^+}=-0.6$.}
\end{figure}

In Fig.~\ref{fig:At12phiM1phimu} we show the contour lines of the CP
asymmetry $A_T$ 
in the $\phi_{M_1}$-$\phi_\mu$ plane
in scenario A (Table~\ref{scentab}) for $\sqrt{s}=500$~GeV and two
sets of beam polarisations $P_{e^-}=-0.8$, $P_{e^+}=+0.6$ and
$P_{e^-}=+0.8$, $P_{e^+}=-0.6$.
In the scenario considered
the $\phi_{M_1}$ dependence is stronger than that on $\phi_\mu$.
In the case of $P_{e^-}=-0.8$, $P_{e^+}=+0.6$ the largest
asymmetries $|A_T| \approx 10\,\%$ are reached 
for $\phi_{M_1} \approx 0.2\pi,1.8\pi$, rather independent of $\phi_\mu$
(Fig.~\ref{fig:At12phiM1phimu} (a)).
For $P_{e^-}=+0.8$, $P_{e^+}=-0.6$ (Fig.~\ref{fig:At12phiM1phimu} (b))
the maximal values of $|A_T| \approx 7.5\,\%$ are attained near
$\phi_{M_1} \approx 0.3\pi, \phi_\mu \approx 0.4\pi$ and
$\phi_{M_1} \approx 1.7\pi, \phi_\mu \approx 1.6\pi$.
The sign flip of $A_T$ between Fig.~\ref{fig:At12phiM1phimu} (a) and
(b) and the weak $\phi_\mu$ dependence
follows from the structure 
of the $\tilde{\chi}^0_i \tilde{\ell}_{L,R} \ell$ couplings
and the facts that $m_{\tilde{\ell}_R} < m_{\tilde{\ell}_L}$
and that the magnitude of the
$\tilde{B}$ component of $\tilde{\chi}^0_2$
depends rather strongly on $\phi_{M_1}$.
For right-handed (left-handed) electrons
the $\tilde{B}$ component
($\tilde{B}$ and $\tilde{W}^3$ components)
of $\tilde{\chi}^0_1$ and $\tilde{\chi}^0_2$
contribute, which leads to a more pronounced
$\phi_{M_1}$ dependence than that on  $\phi_\mu$.

\begin{figure}[p]
\begin{picture}(16,17)

\put(2.6,7.3){(c)}
\put(10.7,7.3){(d)}

\put(0.1,8.1){$M_2/GeV$}
\put(8.3,8.1){$M_2/GeV$}
\put(6.2,0.1){$|\mu|/GeV$}
\put(14.4,0.1){$|\mu|/GeV$}

\put(3.9,8.05){$\sigma/$fb}
\put(12,8.05){$\sigma/$fb}

\put(7.16,2.7){\tiny 40}
\put(6.4,2.7){\tiny 30}
\put(5.85,2.7){\tiny 20}
\put(5.26,2.7){\tiny 10}
\put(5.13,3.3){\tiny 5}
\put(3.9,3.7){\tiny 2.5}
\put(3.7,4.3){\tiny 5}
\put(3.25,5.15){\tiny 10}
\put(2.85,6.55){\tiny 20}

\put(3.65,5){\tiny A}
\put(5.37,3.6){\tiny B}
\put(0,0.4){\epsfig{file=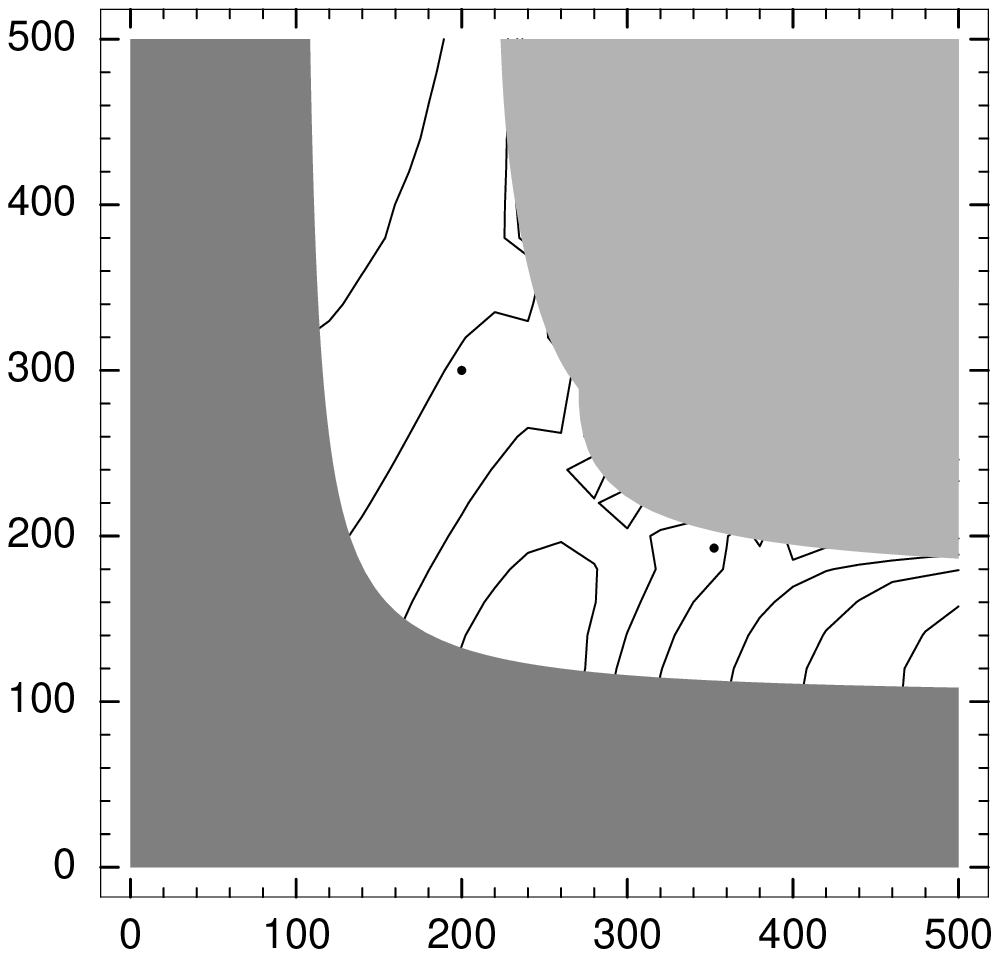,scale=0.77}}

\put(10.75,6.7){\tiny 15}
\put(11,4){\tiny 10}
\put(11.28,3.6){\tiny 5}
\put(11.7,3.2){\tiny 1}
\put(13.9,2.8){\tiny 1}

\put(11.75,5.13){\tiny A}
\put(13.85,3.55){\tiny B}
\put(8.2,0.4){\epsfig{file=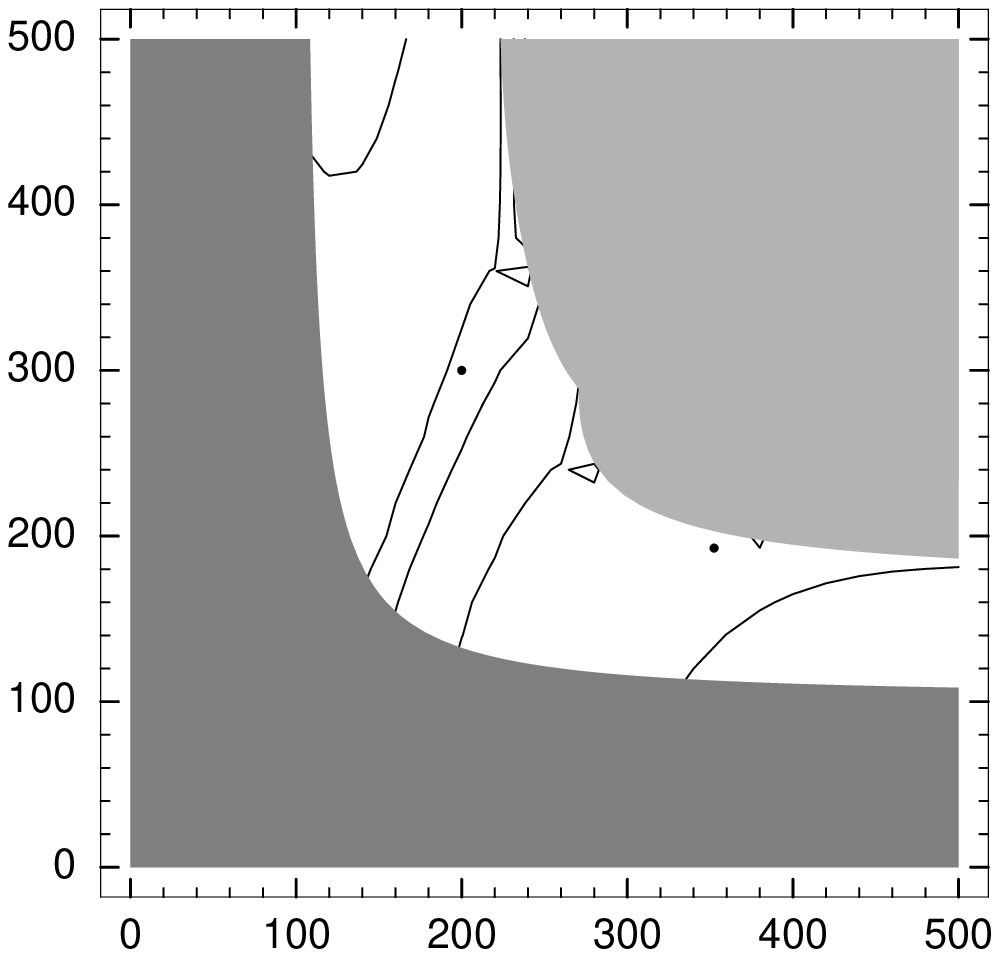,scale=0.77}}

\put(0.1,16.6){$M_2/GeV$}
\put(8.3,16.6){$M_2/GeV$}
\put(6.2,8.6){$|\mu|/GeV$}
\put(14.4,8.6){$|\mu|/GeV$}

\put(3.6,16.55){$A_T$ in \%}
\put(11.8,16.55){$A_T$ in \%}

\put(2.6,15.8){(a)}
\put(10.8,15.8){(b)}

\put(3.75,14.25){\tiny 7.5}
\put(3.2,12.6){\tiny 5}
\put(2.9,12){\tiny 2.5}\put(4.2,11.6){\tiny 2.5}
\put(3.7,11.8){\tiny 1}\put(5.3,12){\tiny 1}\put(2.61,13){\tiny 1}
\put(5.6,11.5){\tiny $-1$}
\put(6.6,11.2){\tiny $-2.5$}
\put(3.65,13.5){\tiny A}
\put(5.65,12.05){\tiny B}
\put(0,8.9){\epsfig{file=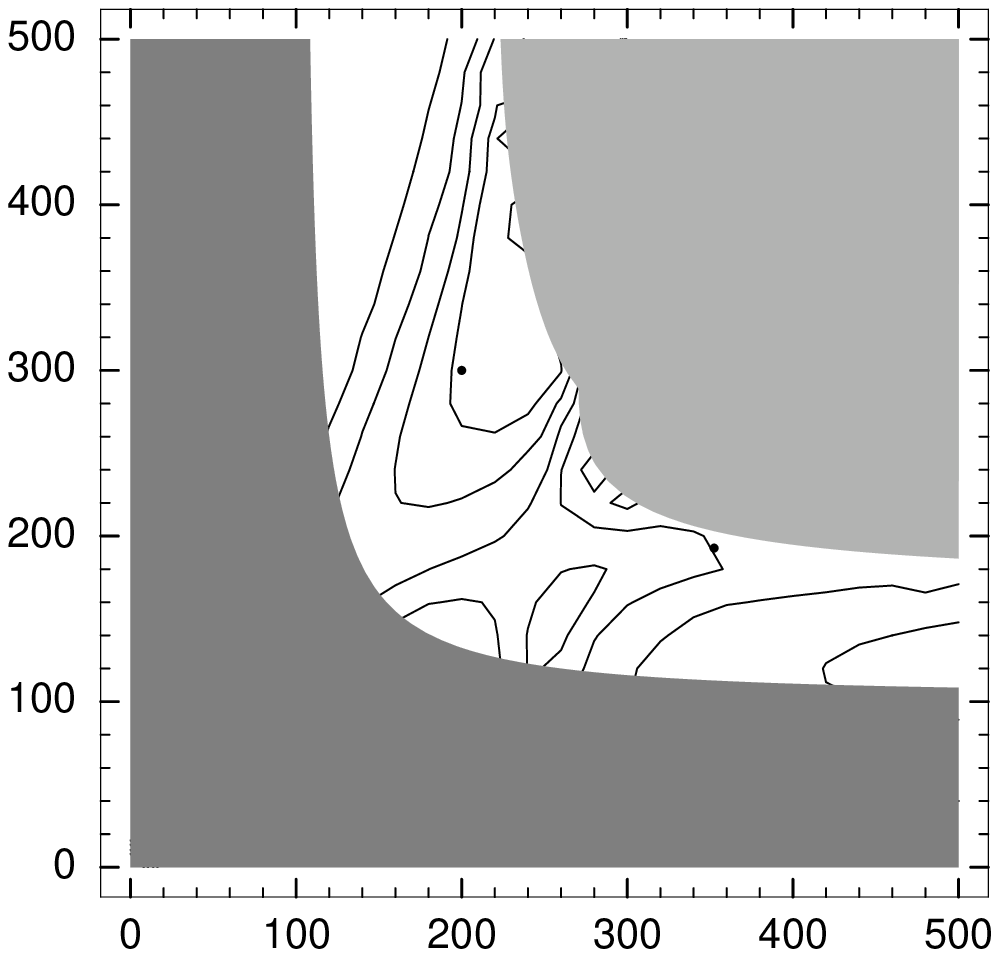,scale=0.77}}

\put(11.9,14.1){\tiny $-10$}
\put(11.9,13.6){\tiny $-7.5$}
\put(11.7,13.2){\tiny $-5$}
\put(11.2,12.6){\tiny $-2.5$}
\put(11.1,12.1){\tiny $-1$}
\put(12.35,11.9){\tiny 1}
\put(11.7,13.63){\tiny A}
\put(13.85,12.05){\tiny B}
\put(8.2,8.9){\epsfig{file=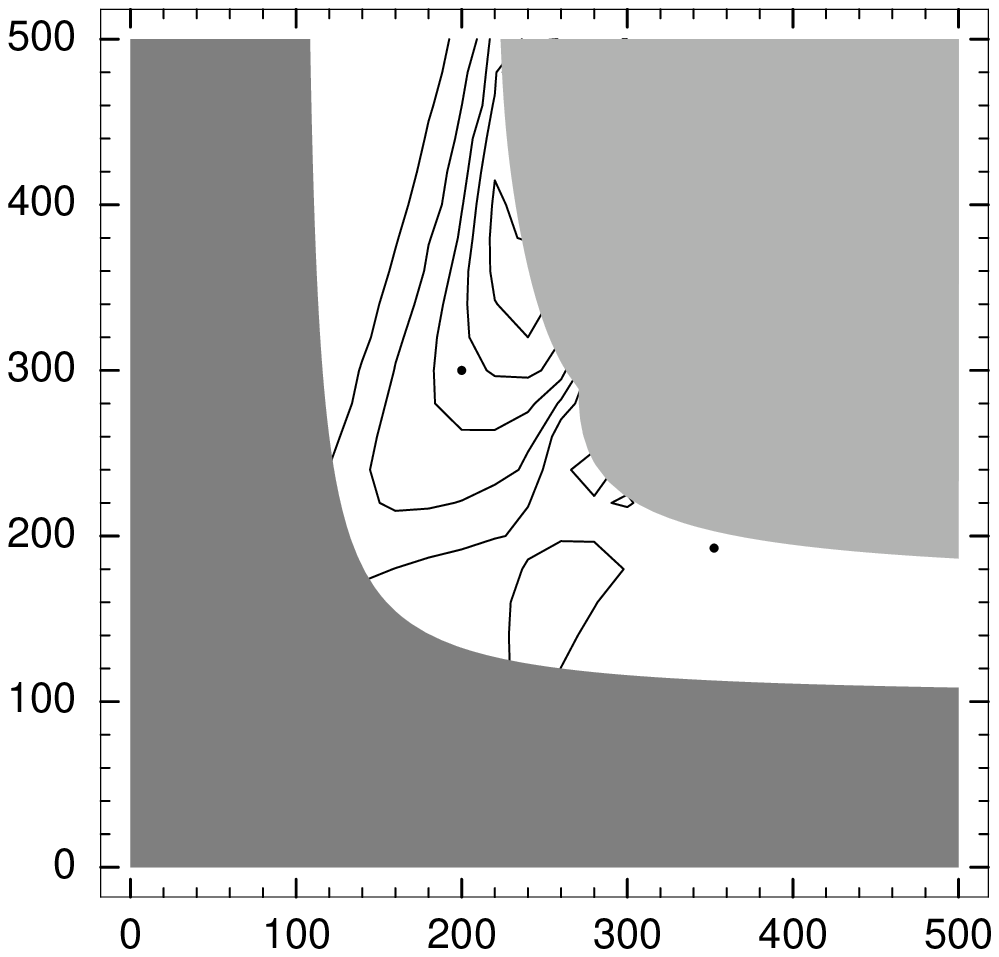,scale=0.77}}

\put(4.5,6.5){\small 2-body decays}
\put(12.7,6.5){\small 2-body decays}
\put(4.5,15){\small 2-body decays}
\put(12.7,15){\small 2-body decays}

\end{picture}
\caption{\label{fig:At12M2mu}
(a), (b) Contours of the CP asymmetry $A_T$ (Eq.~(\ref{Asy})) in \%
for $e^+e^-\rightarrow\tilde{\chi}^0_1\tilde{\chi}^0_2$
with subsequent decay
$\tilde{\chi}^0_2\rightarrow\tilde{\chi}^0_1\ell^+\ell^-$
and (c), (d) contours of the corresponding cross section 
$\sigma(e^+e^- \to \tilde{\chi}^0_1\tilde{\chi}^0_2)\cdot
BR(\tilde{\chi}^0_2\rightarrow\tilde{\chi}^0_1 \ell^+\ell^-)$,
summed over $\ell=e,\mu\,$,
in fb, respectively,
for $\tan\beta = 10$, $m_{\tilde{\ell}_L} = 267.6$~GeV,
$m_{\tilde{\ell}_R} = 224.4$~GeV, $|M_1|/M_2 = 5/3 \tan^2\theta_W$,
$\phi_{M_1}=0.5\pi$ and $\phi_{\mu}=0$
with $\sqrt{s}=500$~GeV and (a), (c) $P_{e^-}=-0.8$, $P_{e^+}=+0.6$ and
(b), (d) $P_{e^-}=+0.8$, $P_{e^+}=-0.6$.
The points mark scenarios A and B of Table~\ref{scentab}.
The dark shaded area marks the parameter space with
$m_{\tilde{\chi}^\pm_1} < 103.5$~GeV excluded by LEP.
In the light shaded area the analysed three-body decay is strongly
suppressed because $m_{\tilde{\chi}^0_2} > m_Z + m_{\tilde{\chi}^0_1}$
or $m_{\tilde{\chi}^0_2} > m_{\tilde{\ell}_R}$.
}
\end{figure}

In Fig.~\ref{fig:At12M2mu} (a) -- (d) we show the contour lines of the
CP asymmetry $A_T$ and of the cross section
for $e^+e^- \to \tilde{\chi}^0_1\tilde{\chi}^0_2$,
$\tilde{\chi}^0_2 \to \tilde{\chi}^0_1 \ell^+ \ell^-$,
summed over $\ell=e,\mu$,
in the $M_2$-$|\mu|$ plane for $|M_1|/M_2 = 5/3 \tan^2\theta_W$,
$\phi_{M_1}=0.5\pi$ and $\phi_{\mu}=0$,
$\tan\beta = 10$, $m_{\tilde{\ell}_L} = 267.6$~GeV,
$m_{\tilde{\ell}_R} = 224.4$~GeV,  
at $\sqrt{s}=500$~GeV and two
sets of beam polarisations, $P_{e^-}=-0.8$, $P_{e^+}=+0.6$ and
$P_{e^-}=+0.8$, $P_{e^+}=-0.6$.
For both polarisation configurations
$A_T$ is largest in the region $|\mu| \approx 240$~GeV and
$M_2 \gtrsim 300$~GeV with maximal values $A_T \approx 7.5\,\%$ for
$P_{e^-}=-0.8$, $P_{e^+}=+0.6$ (Fig.~\ref{fig:At12M2mu} (a))
and $A_T \approx -10\,\%$ for $P_{e^-}=+0.8$, $P_{e^+}=-0.6$
(Fig.~\ref{fig:At12M2mu} (b)).
For $P_{e^-} = -0.8$ ($P_{e^-} = +0.8$)
the main contributions to the asymmetry $A_T$ come from
the $Z$-$\tilde{\ell}_L$ ($Z$-$\tilde{\ell}_R$) interference terms,
therefore large asymmetries are attained if
both neutralinos have significant higgsino
and gaugino components. 
In Fig.~\ref{fig:At12M2mu} (c) and (d) we show the cross
section $\sigma(e^+e^- \to \tilde{\chi}^0_1\tilde{\chi}^0_2
\to \tilde{\chi}^0_1\tilde{\chi}^0_1 \ell^+ \ell^-)$, summed over $\ell=e,\mu$.
For $P_{e^-}=-0.8$, $P_{e^+}=+0.6$
(Fig.~\ref{fig:At12M2mu} (c)) the cross section is larger than about
10~fb in the region with maximal $A_T$ ($|\mu| \approx 240$~GeV,
$M_2 \gtrsim 300$~GeV). For $|\mu| \gtrsim 400$~GeV and
$M_2 \approx 125$~GeV 
the cross section reaches values larger than 40~fb, the asymmetry,
however, is rather small, $A_T \approx -2.5\,\%$.
For $P_{e^-}=+0.8$, $P_{e^+}=-0.6$ (Fig.~\ref{fig:At12M2mu} (d)) the
cross section has values between 1 and 10~fb in the region where $A_T$
is maximal ($|\mu| \approx 240$~GeV, $M_2 \gtrsim 300$~GeV).

In principle, also for the case of hadronic neutralino decays
into heavy quarks,
$\tilde{\chi}^0_2 \to \tilde{\chi}^0_1 b \bar{b}$ and
$\tilde{\chi}^0_2 \to \tilde{\chi}^0_1 c \bar{c}$,
it will be possible to measure the asymmetry $A_T$.
We have calculated $A_T$ in the $M_2$-$|\mu|$ plane and found values 
of roughly the same order of magnitude as those shown in
Fig.~\ref{fig:At12M2mu} for the leptonic decay.
The cross sections are larger by roughly a factor 5 -- 10
in the main part of the parameter region of 
Fig.~\ref{fig:At12M2mu} and approximately of the same order
of magnitude as in the leptonic case for $|\mu| \gtrsim 400$~GeV.
The experimental errors, however, are expected to be
larger than for the leptonic $\tilde{\chi}^0_2$ decays,
because the distinction of the
$b$ and $\bar{b}$ or $c$ and $\bar{c}$ charges 
and the resolution of the jet momenta are
worse than in the leptonic case \cite{quarktagging}.

\begin{figure}[ht!]
\begin{picture}(16,11.5)

\put(0.2,11.2){$A_T$ in \%}
\put(8.4,11.2){$A_T$ in \%}
\put(15.1,5.8){$\phi_{\mu}/\pi$}
\put(6.7,5.8){$\phi_{M_1}/\pi$}
\put(0.9,10.45){(a)}
\put(9.1,10.45){(b)}

\put(0,6.1){\epsfig{file=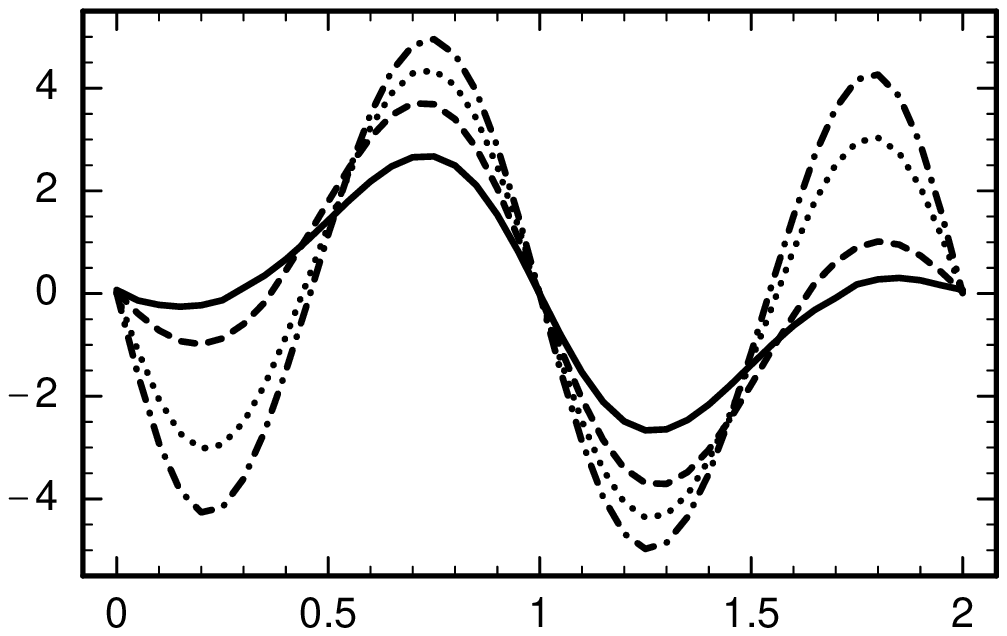,scale=0.77}}
\put(8.15,6.1){\epsfig{file=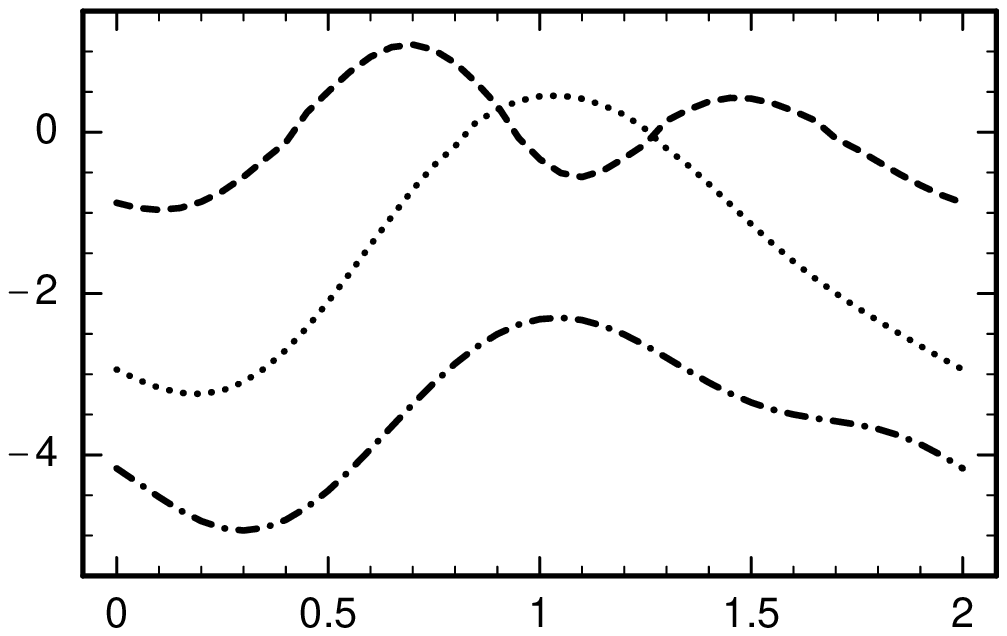,scale=0.77}}

\put(0.2,5.45){$\sigma/$fb}
\put(8.4,5.45){$\sigma/$fb}
\put(15.1,0.1){$\phi_{\mu}/\pi$}
\put(6.7,0.1){$\phi_{M_1}/\pi$}
\put(0.9,4.8){(c)}
\put(9.1,4.8){(d)}

\put(0.05,0.4){\epsfig{file=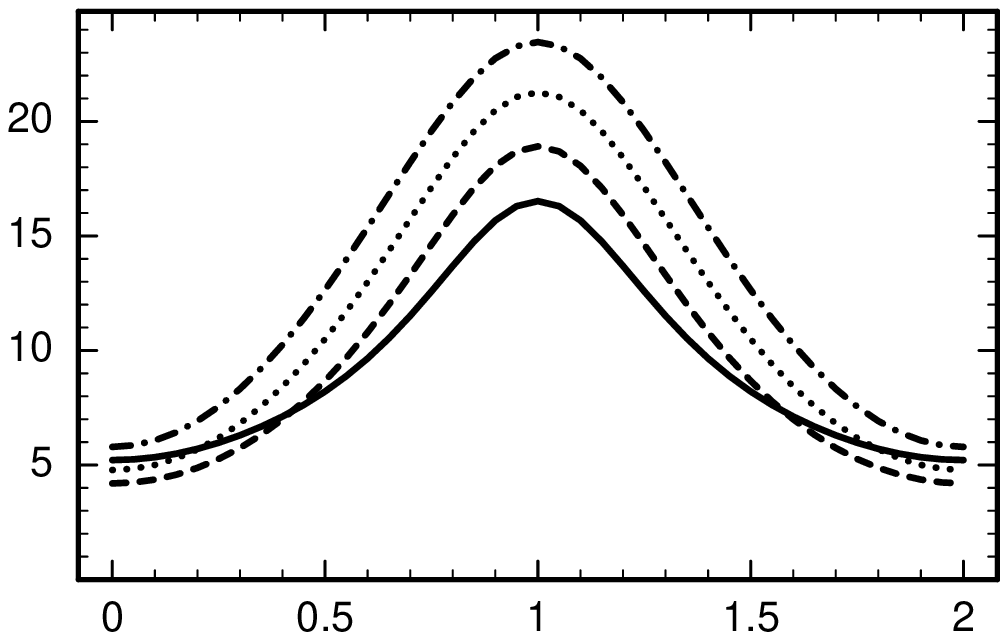,scale=0.765}}
\put(8.2,0.4){\epsfig{file=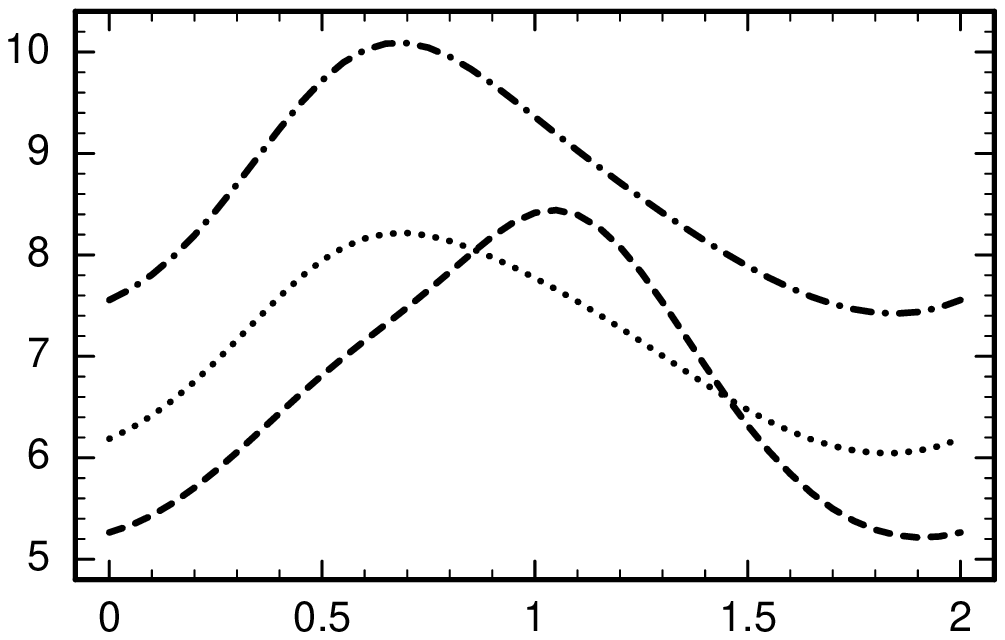,scale=0.765}}

\end{picture}
\caption{\label{fig:AtSPS1a12}
(a), (b) CP asymmetry $A_T$ (Eq.~(\ref{Asy})) and (c), (d) cross section
$\sigma(e^+e^- \to \tilde{\chi}^0_1\tilde{\chi}^0_2)\cdot
BR(\tilde{\chi}^0_2\rightarrow\tilde{\chi}^0_1 \ell^+\ell^-)$,
summed over $\ell=e,\mu\,$,
in scenario B of Table~\ref{scentab} 
with (a), (c) $\phi_\mu = 0$, (b), (d)  $\phi_{M_1} = 0.25\pi$
and $|M_1| = 90$~GeV (solid),
$|M_1| = 95$~GeV (dashed), $|M_1| = 100$~GeV (dotted) and
$|M_1| = 105$~GeV (dashdotted)
for $P_{e^-}=-0.8$, $P_{e^+}=+0.6$ and $\sqrt{s}=500$ GeV.}
\end{figure}

We also studied the $|M_1|$ dependence of the asymmetry $A_T$
and the cross section, relaxing the GUT relation
for $|M_1|$ with all other parameters as in scenario B of
Table~\ref{scentab}.
In Fig.~\ref{fig:AtSPS1a12} (a) we show 
the T-odd asymmetry $A_T$
as a function of $\phi_{M_1}$ for four values of $|M_1| =
90$~GeV, 95~GeV, 100~GeV and 105~GeV.
Here maximal asymmetries $|A_T| \approx 4\,\%$ are reached.
For $|M_1| = 105$~GeV $A_T$ shows nearly two complete oscillations,
whereas for $|M_1| = 90$~GeV there is only one oscillation with extrema
at $\phi_{M_1} = 0.75\pi$ and $1.25\pi$.
In both cases this behaviour is caused 
by cancellations between the contributions 
$\Sigma^{a,\mathrm{O}}_P(\tilde{\chi}^0_i)
 \Sigma^{a,\mathrm{E}}_D(\tilde{\chi}^0_i)$
from production and 
$\Sigma^{a,\mathrm{E}}_P(\tilde{\chi}^0_i)
 \Sigma^{a,\mathrm{O}}_D(\tilde{\chi}^0_i)$
from decay in Eq.~(\ref{CPinT}),
which have opposite sign.
For $|M_1| = 90$~GeV the contributions from the decay are dominant,
whereas for $|M_1| = 105$~GeV both contributions are of similar
magnitude, resulting in two oscillations.
The corresponding cross sections are shown in Fig.~\ref{fig:AtSPS1a12} (c).
In Fig.~\ref{fig:AtSPS1a12} (b) we show $A _T$ as a function of
$\phi_\mu$ for $\phi_{M_1} = 0.25\pi$ and
three values of $|M_1| = 95$~GeV, 100~GeV and 105~GeV.
(We do not discuss $|M_1| = 90$~GeV because in this case
 $m_{\tilde{\chi}^0_2} > m_{\tilde{\chi}^0_1} + m_{Z^0}$ for
 $0.5\pi \lesssim \phi_\mu \lesssim 1.7\pi$.)
For $|M_1| = 100$~GeV $A_T$ exhibits the strongest variation with
$\phi_\mu$ with maximum $|A_T| \approx 3\,\%$ for $\phi_\mu \approx
0.2\pi$ and minimum $|A_T| \approx 0$ for $\phi_\mu \approx \pi$.
The corresponding cross sections are shown in Fig.~\ref{fig:AtSPS1a12} (d).

Furthermore we have analysed the dependence of the asymmetry $A_T$
on the masses of the right and left sleptons, assuming
$m_{\tilde{\ell}_R}$ and $m_{\tilde{\ell}_L}$ are free parameters, 
for $\phi_{M_1}=0.5\pi$ and $\phi_{\mu}=0$ with the other parameters as
in scenario A of Table~\ref{scentab}.
For $\sqrt{s}=500$~GeV and 
beam polarisations $P_{e^-}=-0.8$ and $P_{e^+}=+0.6$
we find that
the asymmetry $A_T$ is larger than 6\,\% for 
$m_{\tilde{\ell}_R} \approx 200$~GeV and
$m_{\tilde{\ell}_L} \gtrsim 150$~GeV.
For $m_{\tilde{\ell}_L} \approx 200$~GeV and
$m_{\tilde{\ell}_R} \gtrsim 300$~GeV the asymmetry is smaller than
$-6\,\%$.
This is due to the fact, that for these beam polarisations the
contributions from $\tilde{\ell}_L$ exchange
to the CP asymmetry are dominant.
Outside of these narrow regions around $m_{\tilde{\ell}_R} \approx 200$~GeV
or $m_{\tilde{\ell}_L} \approx 200$~GeV the asymmetry is very small.
For opposite beam polarisations $P_{e^-}=+0.8$, $P_{e^+}=-0.6$
the $\tilde{\ell}_R$ contributions to $A_T$ are dominant and
the sign of the asymmetry changes: for $m_{\tilde{\ell}_R} \approx
200$~GeV and $m_{\tilde{\ell}_L} \gtrsim 250$~GeV it is 
$A_T \lesssim -11\,\%$ 
and for $m_{\tilde{\ell}_L} \approx 200$~GeV and
$300 \lesssim m_{\tilde{\ell}_R} \lesssim 850$~GeV
it is $A_T \gtrsim 6\,\%$.

\subsection{\boldmath $e^+e^- \to \tilde{\chi}^0_2\tilde{\chi}^0_2$}

\begin{figure}[t]
\begin{picture}(16,8.5)

\put(2.6,7.3){(a)}
\put(10.8,7.3){(b)}

\put(0.1,8.1){$M_2/GeV$}
\put(8.3,8.1){$M_2/GeV$}
\put(6.2,0.1){$|\mu|/GeV$}
\put(14.4,0.1){$|\mu|/GeV$}

\put(3.6,8.05){$A_T$ in \%}
\put(2.65,5.3){\tiny $0.5$}
\put(3.78,3.45){\tiny $1$}\put(3.03,4.45){\tiny $1$}
\put(3.35,3.9){\tiny $1.5$}
\put(5,2.8){\tiny $1.5$}
\put(4.58,2.8){\tiny $1$}
\put(4.02,3){\tiny $0.5$}
\put(6.48,3.07){\tiny $1.75$}
\put(0,0.4){\epsfig{file=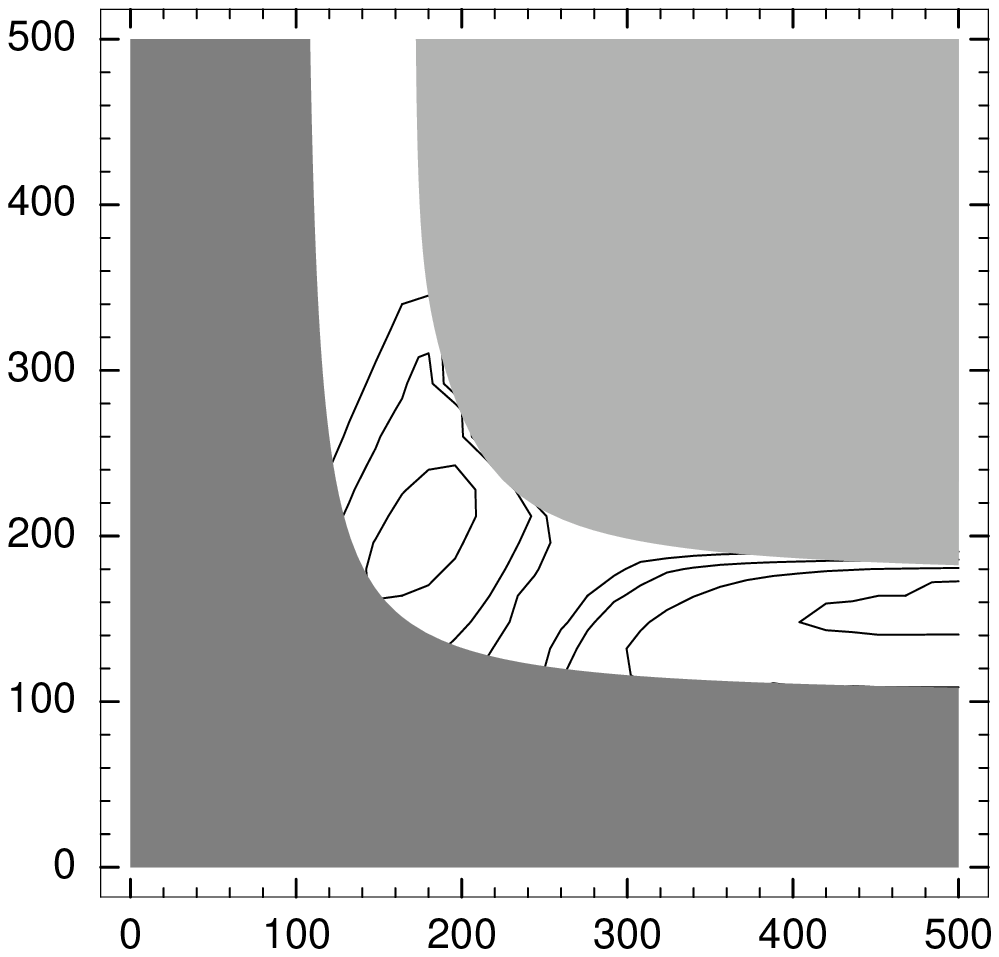,scale=0.77}}

\put(12,8.05){$\sigma$ in fb}

\put(12.3,3.2){\tiny 0.5}
\put(11.85,3.1){\tiny 1}
\put(12.65,2.9){\tiny 1}
\put(13.17,3.05){\tiny 2.5}
\put(13.3,2.77){\tiny 5}
\put(13.7,2.7){\tiny 10}
\put(14.4,2.6){\tiny 15}

\put(8.2,0.4){\epsfig{file=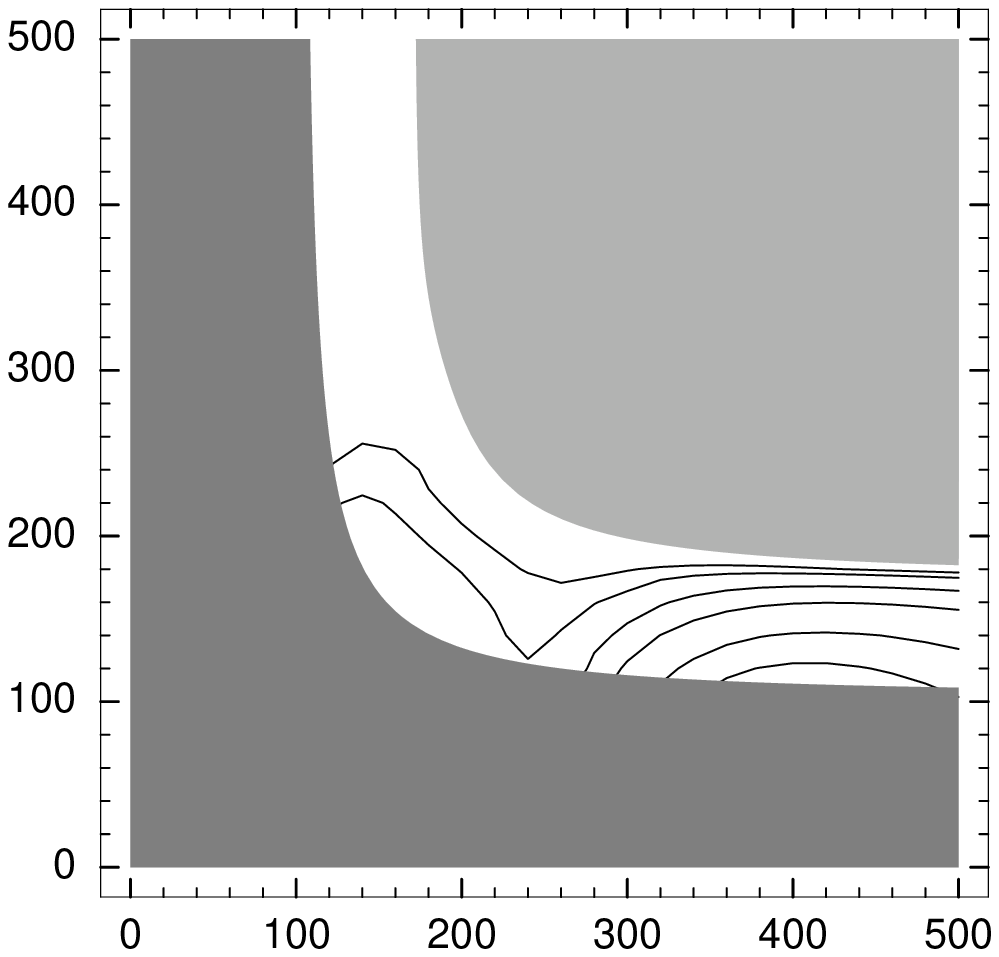,scale=0.77}}

\put(4.5,6.5){\small $2 m_{\tilde{\chi}^0_2} > \sqrt{s}$}
\put(12.7,6.5){\small $2 m_{\tilde{\chi}^0_2} > \sqrt{s}$}

\end{picture}
\caption{\label{fig:At22M2mu350}
Contours (a) of the CP asymmetry $A_T$ (Eq.~(\ref{Asy})) in \%
for $e^+e^-\rightarrow\tilde{\chi}^0_2\tilde{\chi}^0_2$
with subsequent decay
$\tilde{\chi}^0_2\rightarrow\tilde{\chi}^0_1\ell^+\ell^-$
and (b) of the corresponding cross section
$\sigma(e^+e^- \to \tilde{\chi}^0_2\tilde{\chi}^0_2)\cdot
BR(\tilde{\chi}^0_2\rightarrow\tilde{\chi}^0_1 \ell^+\ell^-)\cdot
BR(\tilde{\chi}^0_2\rightarrow\tilde{\chi}^0_1 q\bar{q})$,
summed over $\ell=e,\mu\,$ and $q=u,d,s,c,b$, in fb,
respectively,
for $\tan\beta = 10$, $m_{\tilde{\ell}_L} = 267.6$~GeV,
$m_{\tilde{\ell}_R} = 224.4$~GeV, $|M_1|/M_2 = 5/3 \tan^2\theta_W$,
$\phi_{M_1}=0.5\pi$ and $\phi_{\mu}=0$
with $\sqrt{s}=350$~GeV and $P_{e^-}=-0.8$, $P_{e^+}=+0.6$.
The dark shaded area marks the parameter space with
$m_{\tilde{\chi}^\pm_1} < 103.5$~GeV excluded by LEP.
The light shaded area is kinematically not accessible.
}
\end{figure}

In the reaction $e^+e^- \to \tilde{\chi}^0_2\tilde{\chi}^0_2$
there are only T-odd contributions from the decay, i.e.\
from the second term in Eq.~(\ref{CPinT}).
There is no T-odd contribution from the production,
because the amplitudes Eqs.~(\ref{eq_a1}) -- (\ref{eq_a3})
are real and the first term in Eq.~(\ref{CPinT}) vanishes.
We will give numerical results
for the case where one neutralino decays leptonically
and the other decays hadronically.
If both neutralinos would decay into lepton pairs
of the same flavour,
it may be experimentally difficult to distinguish
the two lepton pairs coming from different
neutralinos.
In Fig.~\ref{fig:At22M2mu350} we show the contour lines 
in the $M_2$-$|\mu|$ plane
of the CP asymmetry $A_T$ and of the cross section
with one neutralino decaying leptonically, 
$\tilde{\chi}^0_2\rightarrow\tilde{\chi}^0_1 \ell^+ \ell^-$, 
summed over $\ell=e,\mu$,
and the other neutralino decaying hadronically,
$\tilde{\chi}^0_2\rightarrow\tilde{\chi}^0_1 q\bar{q}$,
summed over $q=u,d,s,c,b$,
for $\sqrt{s}=350$~GeV and $P_{e^-}=-0.8$, $P_{e^+}=+0.6$. 
The other parameters are $\tan\beta = 10$, $m_{\tilde{\ell}_L} = 267.6$~GeV,
$m_{\tilde{\ell}_R} = 224.4$~GeV, $|M_1|/M_2 = 5/3 \tan^2\theta_W$,
$\phi_{M_1}=0.5\pi$ and $\phi_{\mu}=0$.
For this process $A_T$ is considerably smaller than for
$e^+e^-\rightarrow\tilde{\chi}^0_1\tilde{\chi}^0_2$,
because for production of a pair of equal neutralinos only the decay
amplitudes contribute to the asymmetry.
The largest asymmetries are obtained for $|\mu| \gtrsim 400$~GeV and
$M_2 \approx 150$~GeV, where $\tilde{\chi}^0_1$ and $\tilde{\chi}^0_2$
have gaugino character. The maximum values are $A_T \approx 1.8\,\%$.
The cross section, however, reaches values up to 10~fb in this region.
The hadronic branching ratio 
$\sum_{q=u,d,s,c,b} BR(\tilde{\chi}^0_2\rightarrow\tilde{\chi}^0_1 q\bar{q})$ 
is between 50\,\% and 95\,\% in the main part of the parameter region of
Fig.~\ref{fig:At22M2mu350} and between 30\,\% and 50\,\% for
$|\mu| \gtrsim 400$~GeV.
For $\sqrt{s}=500$~GeV the asymmetries are smaller with maximum of 0.7\,\%
for $|\mu| \gtrsim 400$~GeV and $M_2 \approx 150$~GeV.

We want to remark that it may also be possible to distinguish the two
lepton pairs, if one neutralino decays to an
$e^+e^-$ pair, the other one to a $\mu^+\mu^-$ pair.
Similarly it may be possible to distinguish the two quark pairs,
if one neutralino decays to
a $b\bar{b}$ pair and the other one to a $c\bar{c}$ pair 
\cite{quarktagging}
or pair of light quarks.
However, in the case of both neutralinos decaying into lepton pairs
the event rate is expected to be very low, whereas in the case of the
decay into quark pairs of different flavours the experimental errors
are expected to be much larger.

\subsection{\boldmath $e^+e^- \to \tilde{\chi}^0_3\tilde{\chi}^0_2$}

\begin{figure}[ht!]
\begin{picture}(16,8.6)

\put(0.2,8.3){$\phi_{M_1}/\pi$}
\put(8.4,8.3){$\phi_{M_1}/\pi$}
\put(6.9,0.1){$\phi_{\mu}/\pi$}
\put(15.1,0.1){$\phi_{\mu}/\pi$}

\put(0.9,7){(a)}
\put(9.1,7){(b)}

\put(4.05,7.75){\tiny 0}
\put(4.15,7.44){\tiny $-5$}
\put(4.05,7.1){\tiny $-9$}
\put(4.15,5.75){\tiny $-5$}
\put(4.3,4.3){\tiny 0}
\put(4.3,3.05){\tiny 5}
\put(4.3,1.7){\tiny 9}
\put(4.3,1.38){\tiny 5}
\put(4.3,1.1){\tiny 0}

\put(0,0.4){\epsfig{file=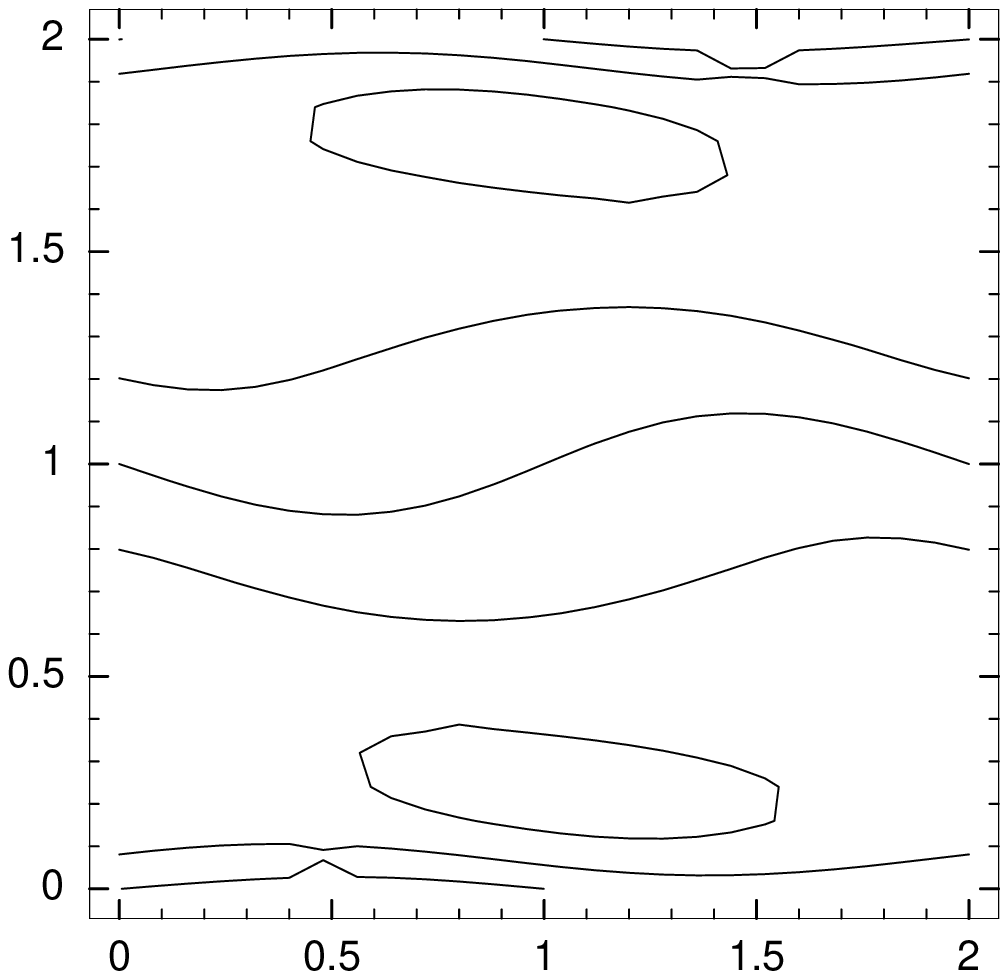,scale=0.77}}

\put(12.3,7.75){\tiny 0}
\put(14.5,7.2){\tiny 3}
\put(13.75,6.95){\tiny 5}
\put(12.95,6.7){\tiny 7}
\put(12.5,5.37){\tiny 3}
\put(12.5,4.3){\tiny 0}
\put(12.3,3.5){\tiny $-3$}
\put(11.7,2.15){\tiny $-7$}
\put(10.95,1.85){\tiny $-5$}
\put(10.2,1.65){\tiny $-3$}
\put(12.5,1.1){\tiny 0}

\put(8.2,0.4){\epsfig{file=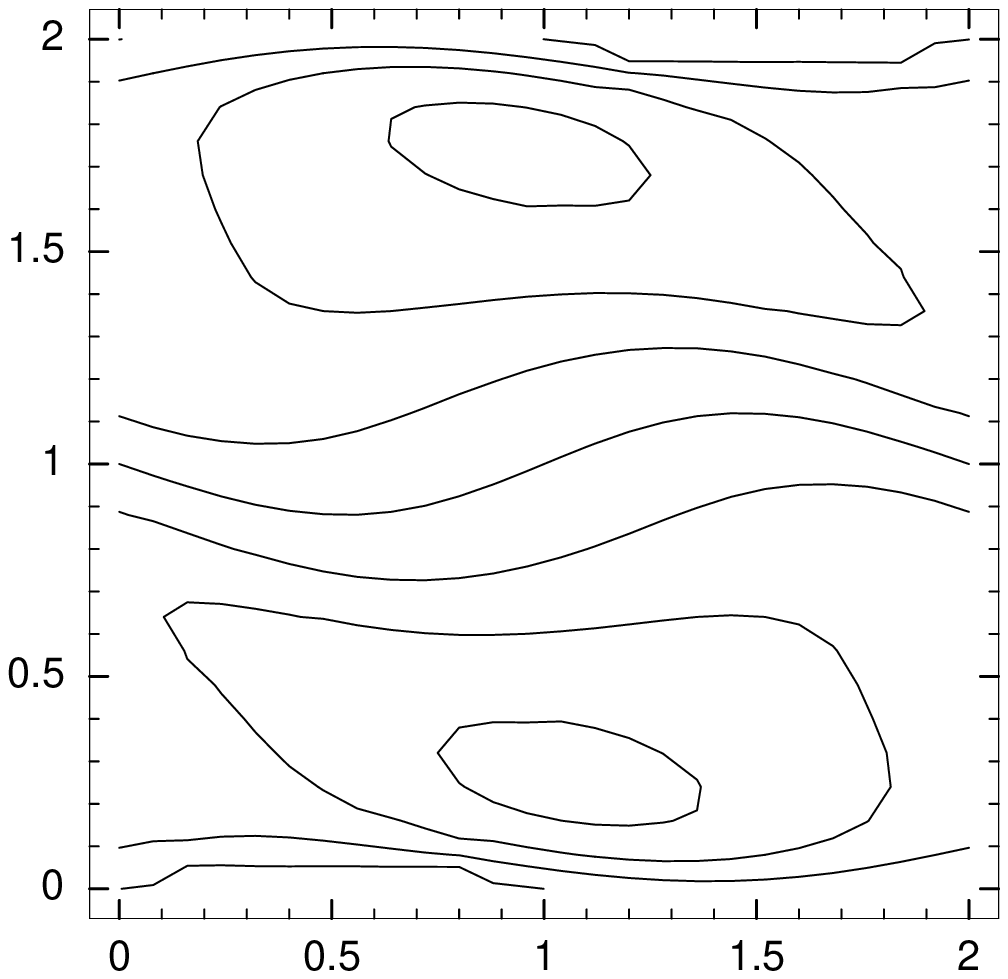,scale=0.77}}

\end{picture}
\caption{\label{fig:At32phiM1phimu}
Contours of the CP asymmetry $A_T$ (Eq.~(\ref{Asy})) in \%
for $e^+e^- \to \tilde{\chi}^0_3\tilde{\chi}^0_2$
with subsequent decay
$\tilde{\chi}^0_2 \to \tilde{\chi}^0_1\ell^+\ell^-$
in scenario A of Table~\ref{scentab} for
$\sqrt{s}=500$~GeV and (a) $P_{e^-}=-0.8$, $P_{e^+}=+0.6$ and
(b) $P_{e^-}=+0.8$, $P_{e^+}=-0.6$.}
\end{figure}

In Fig.~\ref{fig:At32phiM1phimu} we show the contour lines of the CP
asymmetry $A_T$ in the $\phi_{M_1}$-$\phi_\mu$ plane
in scenario A (Table~\ref{scentab}) for $\sqrt{s}=500$~GeV and two
sets of beam polarisations $P_{e^-}=-0.8$, $P_{e^+}=+0.6$ and
$P_{e^-}=+0.8$, $P_{e^+}=-0.6$.
As in the case of $\tilde{\chi}^0_1\tilde{\chi}^0_2$ production
(Fig.~\ref{fig:At12phiM1phimu})
the $\phi_{M_1}$ dependence is stronger than that on $\phi_\mu$ in
this scenario, since the $\tilde{B}$ component of the decaying
particle $\tilde{\chi}^0_2$ depends stronger on $\phi_{M_1}$. The largest
asymmetries of $|A_T| \gtrsim 9\,\%$ are reached in the case of
$P_{e^-}=-0.8$, $P_{e^+}=+0.6$ (Fig.~\ref{fig:At32phiM1phimu} (a))
for $\phi_{M_1} \approx 0.25\pi,1.75\pi$ and
$0.6\pi \lesssim \phi_\mu \lesssim 1.4\pi$.
In the case of $P_{e^-}=+0.8$, $P_{e^+}=-0.6$ 
(Fig.~\ref{fig:At32phiM1phimu} (b)) the sign of $A_T$ flips and the
largest asymmetries $|A_T| \gtrsim 7\,\%$ appear for 
$\phi_{M_1} \approx 0.25\pi,1.75\pi$ and
$0.8\pi \lesssim \phi_\mu \lesssim 1.2\pi$.

\begin{figure}[p]
\begin{picture}(16,17)

\put(2.83,7.3){(c)}
\put(10.8,7.3){(d)}

\put(0.1,8.1){$M_2/GeV$}
\put(8.3,8.1){$M_2/GeV$}
\put(6.2,0.1){$|\mu|/GeV$}
\put(14.4,0.1){$|\mu|/GeV$}

\put(3.9,8.05){$\sigma/$fb}
\put(12,8.05){$\sigma/$fb}

\put(4.9,2.9){\tiny 1}
\put(2.6,6.85){\tiny 1}\put(4.5,3.8){\tiny 1}
\put(3.4,3.1){\tiny 2}\put(2.65,5.95){\tiny 2}
\put(3.1,3.4){\tiny 5}\put(2.68,4.8){\tiny 5}
\put(3,4.28){\tiny 10}
\put(3.75,5.55){\tiny 15}
\put(3.35,5){\tiny A}
\put(0,0.4){\epsfig{file=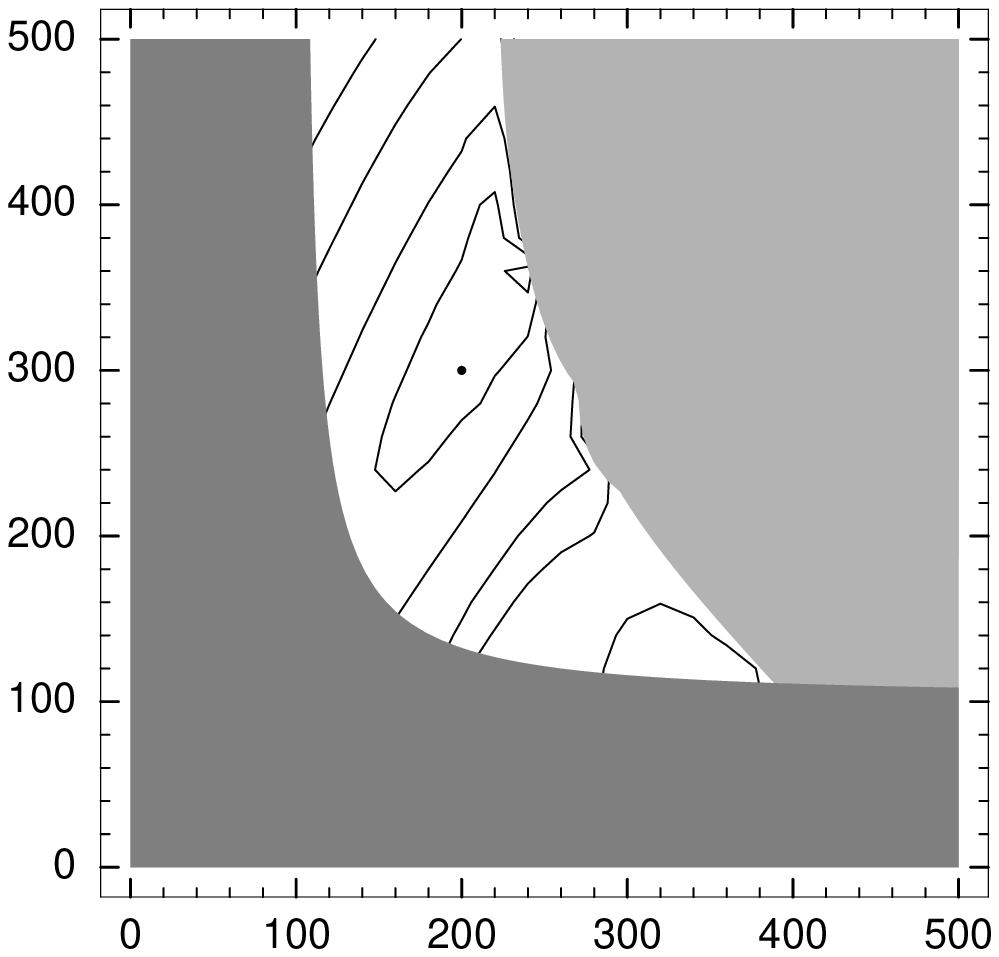,scale=0.77}}

\put(11.58,3.22){\tiny 1}\put(10.8,5.85){\tiny 1}
\put(11.27,3.45){\tiny 2}\put(10.82,5.28){\tiny 2}
\put(11,3.95){\tiny 4}\put(10.87,4.72){\tiny 4}
\put(12.05,5.45){\tiny 6}
\put(11.55,5){\tiny A}
\put(8.2,0.4){\epsfig{file=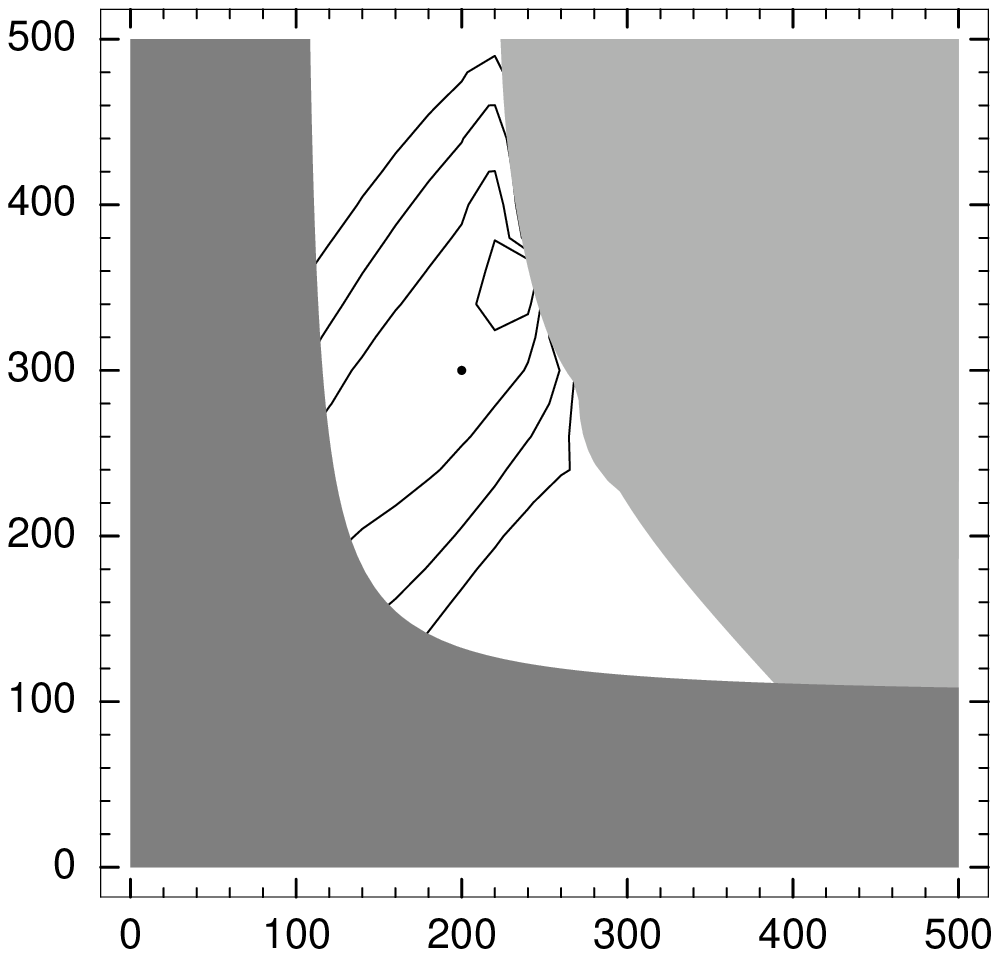,scale=0.77}}

\put(0.1,16.6){$M_2/GeV$}
\put(8.3,16.6){$M_2/GeV$}
\put(6.2,8.6){$|\mu|/GeV$}
\put(14.4,8.6){$|\mu|/GeV$}

\put(3.6,16.55){$A_T$ in \%}
\put(11.8,16.55){$A_T$ in \%}

\put(2.6,15.8){(a)}
\put(10.8,15.8){(b)}

\put(3.4,16.17){\tiny 1}\put(4.25,11.65){\tiny 1}
 \put(4.63,12.3){\tiny 1}
\put(3.57,16.17){\tiny 2.5}\put(4,11.9){\tiny 2.5}
 \put(4.58,11.9){\tiny 2.5}
\put(2.85,12.2){\tiny 5}\put(3.47,11.5){\tiny 5}
 \put(4.55,11.3){\tiny 5}
\put(3.3,12.2){\tiny 7.5}\put(4.78,11.27){\tiny 7.5}
\put(3.95,13.15){\tiny 10}\put(5.4,11.22){\tiny 10}
\put(3.4,13.5){\tiny A}
\put(0,8.9){\epsfig{file=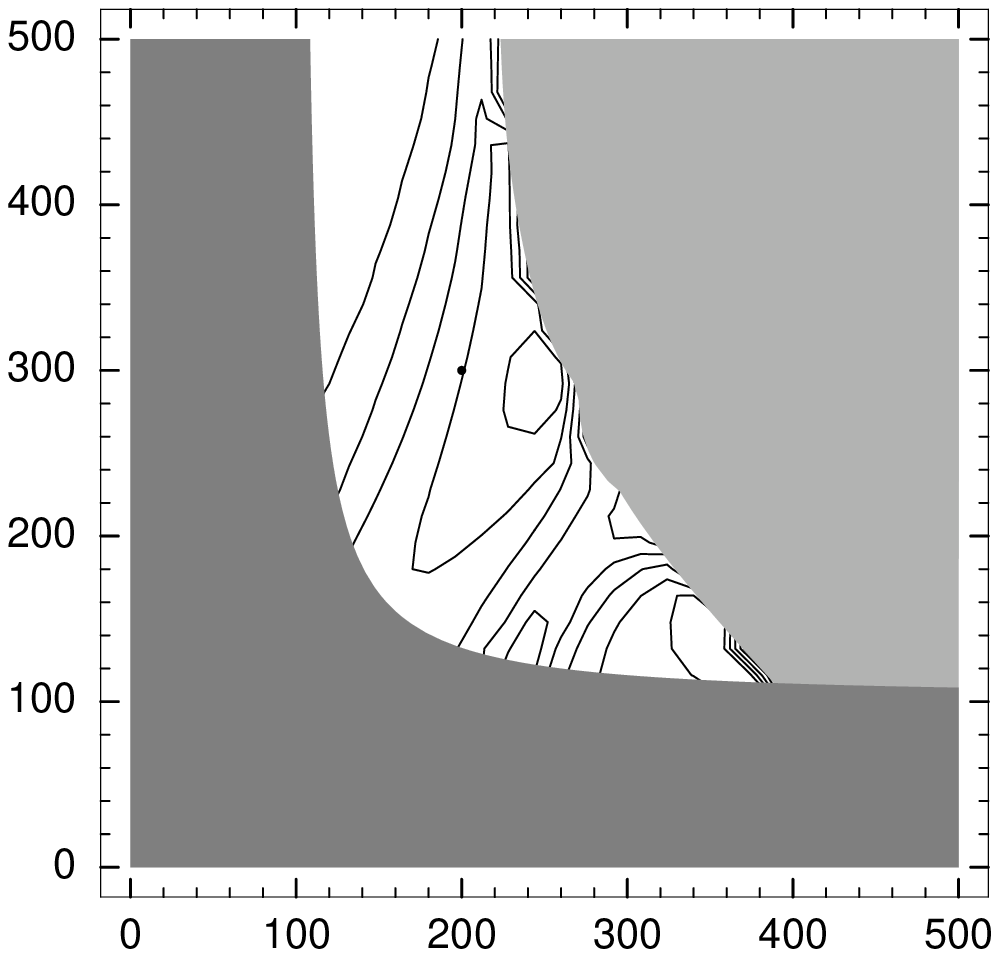,scale=0.77}}

\put(11.73,15.5){\tiny $-1$}
\put(11.9,14.6){\tiny $-3$}\put(11.68,11.6){\tiny $-3$}
\put(11.25,11.8){\tiny $-6$}
\put(12,12.9){\tiny $-8$}
\put(12.3,11.7){\tiny $-1$}\put(12.65,12.15){\tiny $-1$}
\put(12.9,11.8){\tiny $-3$}
\put(13,11.25){\tiny $-6$}
\put(13.48,11.45){\tiny $-8$}
\put(11.6,13.5){\tiny A}
\put(8.2,8.9){\epsfig{file=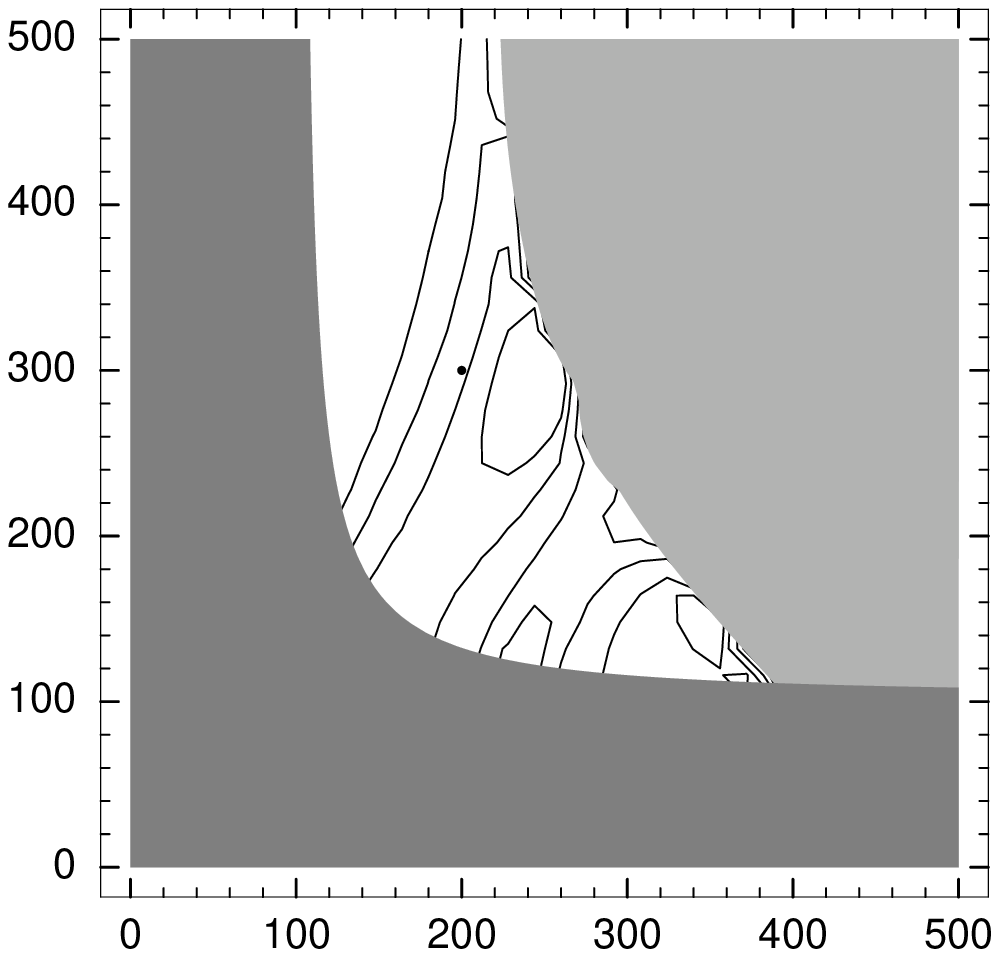,scale=0.77}}

\put(4.3,6.7){\small $m_{\tilde{\chi}^0_2} + m_{\tilde{\chi}^0_3} > \sqrt{s}$}
\put(4.3,6){\small or 2-body decays}
\put(12.5,6.7){\small $m_{\tilde{\chi}^0_2} + m_{\tilde{\chi}^0_3} > \sqrt{s}$}
\put(12.5,6){\small or 2-body decays}
\put(4.3,14.5){\small  or 2-body decays}
\put(4.3,15.2){\small $m_{\tilde{\chi}^0_2} + m_{\tilde{\chi}^0_3} > \sqrt{s}$}
\put(12.5,14.5){\small or 2-body decays}
\put(12.5,15.2){\small $m_{\tilde{\chi}^0_2} + m_{\tilde{\chi}^0_3} > \sqrt{s}$}

\end{picture}
\caption{\label{fig:At32M2mu}
(a), (b) Contours of the CP asymmetry $A_T$ (Eq.~(\ref{Asy})) in \%
for $e^+e^-\rightarrow\tilde{\chi}^0_3\tilde{\chi}^0_2$
with subsequent decay
$\tilde{\chi}^0_2\rightarrow\tilde{\chi}^0_1\ell^+\ell^-$
and (c), (d) contours of the corresponding cross section
$\sigma(e^+e^- \to \tilde{\chi}^0_3\tilde{\chi}^0_2)\cdot
BR(\tilde{\chi}^0_2\rightarrow\tilde{\chi}^0_1 \ell^+\ell^-)$,
summed over $\ell=e,\mu\,$,
in fb, respectively,
for $\tan\beta = 10$, $m_{\tilde{\ell}_L} = 267.6$~GeV,
$m_{\tilde{\ell}_R} = 224.4$~GeV, $|M_1|/M_2 = 5/3 \tan^2\theta_W$,
$\phi_{M_1}=0.5\pi$ and $\phi_{\mu}=0$
with $\sqrt{s}=500$~GeV and (a), (c) $P_{e^-}=-0.8$, $P_{e^+}=+0.6$ and
(b), (d) $P_{e^-}=+0.8$, $P_{e^+}=-0.6$.
The dark shaded area marks the parameter space with
$m_{\tilde{\chi}^\pm_1} < 103.5$~GeV excluded by LEP.
The light shaded area is kinematically not accessible or
in this area the analysed three-body decay is strongly
suppressed because $m_{\tilde{\chi}^0_2} > m_Z + m_{\tilde{\chi}^0_1}$
or $m_{\tilde{\chi}^0_2} > m_{\tilde{\ell}_R}$, respectively.
}
\end{figure}

In Fig.~\ref{fig:At32M2mu} we show the contour lines of the
CP asymmetry $A_T$ and of the cross section
for the process $e^+e^-\rightarrow\tilde{\chi}^0_3\tilde{\chi}^0_2$
with the subsequent decay
$\tilde{\chi}^0_2\rightarrow\tilde{\chi}^0_1 \ell^+\ell^-$,
summed over $\ell = e,\mu$, in the $M_2$-$|\mu|$ plane
for $\tan\beta = 10$, $m_{\tilde{\ell}_L} = 267.6$~GeV,
$m_{\tilde{\ell}_R} = 224.4$~GeV, $|M_1|/M_2 = 5/3 \tan^2\theta_W$,
$\phi_{M_1}=0.5\pi$ and $\phi_{\mu}=0$
for $\sqrt{s}=500$~GeV and two
sets of beam polarisations $P_{e^-}=-0.8$, $P_{e^+}=+0.6$ and
$P_{e^-}=+0.8$, $P_{e^+}=-0.6$.
$|A_T|$ has two maxima in the regions 
$|\mu| \approx 240$~GeV, $M_2 \approx 300$~GeV and
$|\mu| \approx 350$~GeV, $M_2 \approx 140$~GeV, respectively,
for both polarisation configurations.
The maximum values
are $A_T \gtrsim 10\,\%$ for
$P_{e^-}=-0.8$, $P_{e^+}=+0.6$ (Fig.~\ref{fig:At32M2mu} (a))
and $A_T \lesssim -8\,\%$ for
$P_{e^-}=-0.8$, $P_{e^+}=+0.6$ (Fig.~\ref{fig:At32M2mu} (b)).
The cross section reaches values larger than 10~fb (6~fb) in the region
with maximal asymmetry around
$|\mu| \approx 240$~GeV, $M_2 \approx 300$~GeV
for $P_{e^-}=-0.8$, $P_{e^+}=+0.6$, Fig.~\ref{fig:At32M2mu} (c)
($P_{e^-}=+0.8$, $P_{e^+}=-0.6$, Fig.~\ref{fig:At32M2mu} (d)).
In the region around $|\mu| \approx 350$~GeV, $M_2 \approx 140$~GeV
the cross section is $\lesssim 1$~fb.

We have also calculated the branching ratio 
$BR(\tilde{\chi}^0_3\rightarrow\tilde{\chi}^0_1 \ell^+\ell^-)$,
$\ell = e$ or $\mu$, of $\tilde{\chi}^0_3$.
In parameter regions, where $\tilde{\chi}^0_3$ decays via a three-body
decay and where it is especially difficult to distinguish the two
lepton pairs coming from different neutralinos,
$BR(\tilde{\chi}^0_3\rightarrow\tilde{\chi}^0_1 \ell^+\ell^-)$
is always smaller than about 7\,\%.

Furthermore we have calculated the CP asymmetry $A_T$ for the the process
$e^+e^-\rightarrow\tilde{\chi}^0_4\tilde{\chi}^0_2$
with the subsequent decay
$\tilde{\chi}^0_2\rightarrow\tilde{\chi}^0_1\ell^+\ell^-$,
$\ell = e,\mu$.
For the associated production of $\tilde{\chi}^0_2$ and
$\tilde{\chi}^0_4$ the accessible parameter space at a linear collider
with $\sqrt{s}=500$~GeV is rather small and is approximately
constrained by $M_2 + |\mu| \lesssim 500$~GeV.
For the same parameters as in Fig.~\ref{fig:At32M2mu} 
we find that
the CP asymmetry has maximum values $|A_T| \approx 6\,\%$.
However, in parameter regions with $|A_T| > 2\,\%$
the cross section
$\sigma(e^+e^- \to \tilde{\chi}^0_4\tilde{\chi}^0_2)\cdot
BR(\tilde{\chi}^0_2\rightarrow\tilde{\chi}^0_1 \ell^+\ell^-)$,
summed over $\ell=e,\mu\,$,
is always smaller than 1~fb.

\section{Conclusions}

In this paper we have studied CP violation in
neutralino production and subsequent three-body decay processes.
As CP-sensitive observable we have chosen 
a T-odd asymmetry based on triple
product correlations between the incoming and outgoing particles. A
non-vanishing
value for the asymmetry indicates directly CP-violating effects.
We have provided compact analytical formulae for the cross section of the
whole process as well as for the T-odd asymmetry for longitudinal beam
polarisations.
It is necessary to include the full spin correlations between
production and decay.
The asymmetry can be directly measured in the experiment without
reconstruction
of the momentum of the decaying neutralino or further final-state analyses.
In our numerical results
we have chosen representative scenarios
and have shown that this T-odd asymmetry
can reach values up to 13\,\%
in some parts of the parameter space of the unconstrained MSSM,
even for small CP-violating phases
as may be indicated by the EDM constraints. This observable will
therefore be an important tool in the search for CP-violating effects
in the neutralino sector and the determination of the phases of the
complex SUSY parameters.

\section*{Acknowledgements}

We thank W.~Porod for providing us a version of SPheno
for complex SUSY parameters, prior to its release,
which has been used to compute the neutralino decay widths.
We thank T.~Kernreiter, O.~Kittel and W.~Majerotto for useful discussions.
This work is supported by the `Fonds zur F\"orderung der
wissenschaftlichen For\-schung' of Austria, FWF Project No.~P16592-N02,
by the European Community's Human Potential Programme
under contract HPRN-CT-2000-00149
and by the Deutsche Forschungsgemeinschaft (DFG) under contract No.\
\mbox{FR~1064/5-2}.

\begin{appendix}
\section{Kinematics}

We choose a coordinate frame in the laboratory system, where
the momenta are
\begin{eqnarray}
p_1 &=& E_b(1,-\sin\Theta,0, \cos\Theta)\label{eq_19a},\\
p_2 &=& E_b(1, \sin\Theta,0,-\cos\Theta),\\
p_3 &=& (E_i,0,0,-q),\\
p_4 &=& (E_j,0,0, q).\label{eq_19}
\end{eqnarray}
$\Theta$ is the scattering angle between the incoming
$e^{-}(p_1)$ beam and the outgoing neutralino
$\tilde{\chi}^0_j(p_4)$, the azimuth $\Phi$ can be chosen equal to
zero.
The energy and the momenta of the neutralinos
$\tilde{\chi}^0_j(p_4)$ and $\tilde{\chi}^0_i(p_3)$ are
\begin{equation}
E_i=\frac{s+m_i^2-m_j^2}{2 \sqrt{s}},\quad
 E_j=\frac{s+m_j^2-m_i^2}{2 \sqrt{s}},\quad 
 q=\frac{\sqrt{\lambda(s,m_i^2,m_j^2)}}{2 \sqrt{s}},\label{eq_20}
\end{equation}
where $m_i, m_j$ are the masses of the neutralinos and $\lambda$
is the kinematical triangle function
$\lambda(x,y,z)=x^2+y^2+z^2-2xy-2xz-2yz$.

The polarisation vectors $s^a_{\mu}(\tilde{\chi}^0_i)$ and
$s^b_{\mu}(\tilde{\chi}^0_j)$ ($a,b=1,2,3$) of the neutralinos
in the laboratory system are
\begin{eqnarray}
s^{1\mu}(\tilde{\chi}^0_{i,j})&=&(0,\mp 1,0,0),\label{eq_21}\\
s^{2\mu}(\tilde{\chi}^0_{i,j})&=&(0,0,1,0),\label{eq_22}\\
s^{3\mu}(\tilde{\chi}^0_{i,j})&=&\frac{1}{m_{i,j}}(q,0,0,\mp
E_{i,j}), \label{eq_23}
\end{eqnarray}
where $s^3(\tilde{\chi}^0_{i,j})$ describes the longitudinal
polarisation, $s^1(\tilde{\chi}^0_{i,j})$ the 
transverse polarisation in the scattering plane, and
$s^2(\tilde{\chi}^0_{i,j})$ the 
transverse polarisation perpendicular to the scattering plane.

\section{Explicit expressions of the amplitude squared}

\subsection{Production}

Here we give the analytic expressions for the production spin density matrix.
The terms
\begin{equation}
P(\tilde{\chi}^0_i \tilde{\chi}^0_j)= P(Z Z)+P(Z \tilde{e}_L)+P(Z
\tilde{e}_R)+ P(\tilde{e}_L \tilde{e}_L)+P(\tilde{e}_R
\tilde{e}_R) \label{eq_sump}
\end{equation}
are independent of the polarisation of the
neutralinos \cite{Moortgat-Pick:1999di,Moortgat-Pick:1999wr},
\begin{eqnarray}
P(ZZ)&=&\frac{g^4}{\cos^4\Theta_W} |\Delta^s(Z)|^2 (R^2_{\ell}
  c_R +L^2_{\ell} c_L) \nonumber\\
 && \times \left\{ |O^{''L}_{ij}|^2(f_1+f_2)- [(Re
  O^{''L}_{ij})^2-(Im O^{''L}_{ij})^2]f_3 \right\}, \label{eq_pzz}\\
P(Z\tilde{e}_L)&=& \frac{g^4}{2\cos^2\Theta_W}L_{\ell} c_L \nonumber\\
 && \times Re\Big\{\Delta^s(Z)\Big[-(\Delta^{t*}(\tilde{e}_L)f^{L*}_{\ell
  i}f^L_{\ell j} O^{''L*}_{ij}+ \Delta^{u*}(\tilde{e}_L)f^{L}_{\ell
  i}f^{L*}_{\ell j} O^{''L}_{ij}) f_3\nonumber\\
 &&\phantom{\times Re\Big\{}+ 2 \Delta^{t*}(\tilde{e}_L) 
    f^{L*}_{\ell i} f^L_{\ell j}
  O^{''L}_{ij} f_1+2 \Delta^{u*}(\tilde{e}_L) f^{L}_{\ell i}
  f^{L*}_{\ell j} O^{''L*}_{ij} f_2\Big]\Big\}, \label{eq_pzel}\\
P(\tilde{e}_L\tilde{e}_L)&=&\frac{g^4}{4} c_L \Big\{|f^L_{\ell
  i}|^2 |f^L_{\ell j}|^2 (|\Delta^t(\tilde{e}_L)|^2
  f_1+|\Delta^u(\tilde{e}_L)|^2 f_2 )\nonumber\\
 &&\phantom{\frac{g^4}{4} c_L \Big\{} -Re\{ (f^{L*}_{\ell i})^2
  (f^{L}_{\ell j})^2
  \Delta^u(\tilde{e}_L) \Delta^{t*}(\tilde{e}_L) f_3\} \Big\}.
  \label{eq_pelel}
\end{eqnarray}
$P(Z\tilde{e}_R)$, $P(\tilde{e}_R \tilde{e}_R)$ are obtained by
the following substitutions in Eqs.~(\ref{eq_pzel}), (\ref{eq_pelel}):
\begin{equation}
 L_\ell \to R_\ell, \quad c_L \to c_R, \quad
 \Delta^{t,u}(\tilde{e}_L)\to \Delta^{t,u}(\tilde{e}_R), \quad
 O^{''L}_{ij}\to O^{''R}_{ij}, \quad
 f_{\ell i,j}^L\to f_{\ell i,j}^R.
\label{eq_psub}
\end{equation}
The beam polarisation weighting factors are given by
\begin{eqnarray}
c_L&=&(1-P_{e^-})(1+P_{e^+}), \label{eq_polL}\\
c_R&=&(1+P_{e^-})(1-P_{e^+}) \label{eq_polR}
\end{eqnarray}
and the kinematical factors in invariant form are
\begin{eqnarray}
f_1 & = & (p_1 p_4)(p_2 p_3), \label{eq_kinp1}\\
f_2 & = & (p_1 p_3)(p_2 p_4), \label{eq_kinp2}\\
f_3 & = & m_i m_j (p_1 p_2). \label{eq_kinp3}
\end{eqnarray}

The terms
\begin{equation}
\Sigma_P^a(\tilde{\chi}^0_i)=
 \Sigma_P^a(ZZ)
+\Sigma_P^a(Z\tilde{e}_L) +\Sigma_P^a(Z\tilde{e}_R)
+\Sigma_P^a(\tilde{e}_L \tilde{e}_L) +\Sigma_P^a(\tilde{e}_R
\tilde{e}_R),\label{eq_sumsp1}
\end{equation}
which depend on the polarisation of the neutralino
$\tilde{\chi}^0_i$ and which are
relevant for the calculation of
the CP asymmetries, are given by
\begin{eqnarray}
\Sigma_P^a(ZZ)&=& \frac{g^4}{\cos^4\Theta_W} |\Delta^s(Z)|^2
    (R^2_{\ell} c_{R}- L^2_{\ell} c_{L}) 
    \Big\{ |O^{''L}_{ij}|^2 (f_2^a-f_1^a) \nonumber\\
  & & \hspace{14mm}
    {} - [(Re O^{''L}_{ij})^2 -(Im O^{''L}_{ij})^2] f_3^a
    + 2 (Re O^{''L}_{ij}) (Im O^{''L}_{ij})  i f_4^a\Big\}, \label{eq_spzz}\\
\Sigma_P^a(Z\tilde{e}_L) &=& \frac{g^4}{2 \cos^2\Theta_W} L_{\ell}
    c_{L}
    \nonumber\\
  & & \times Re\Big\{\Delta^s(Z) \Big[2 f^{L}_{\ell i}f^{L*}_{\ell
    j} O^{''L*}_{ij} \Delta^{u*}(\tilde{e}_L)f_1^a -2 f^{L*}_{\ell
    i}f^{L}_{\ell j} O^{''L}_{ij}
    \Delta^{t*}(\tilde{e}_L)f_2^a \nonumber\\
  & & \phantom{\times Re\Big\{\Delta^s(Z) \Big[}
    +\big[f^{L}_{\ell i}f^{L*}_{\ell j} O^{''L}_{ij}
    \Delta^{u*}(\tilde{e}_L) +f^{L*}_{\ell i}f^{L}_{\ell j}
    O^{''L*}_{ij} \Delta^{t*}(\tilde{e}_L)\big]
    f_3^a\nonumber\\
  & & \phantom{\times Re\Big\{\Delta^s(Z) \Big[}
    +\big[f^{L*}_{\ell i}f^{L}_{\ell j} O^{''L*}_{ij}
    \Delta^{t*}(\tilde{e}_L) -f^{L}_{\ell i}f^{L*}_{\ell j}
    O^{''L}_{ij} \Delta^{u*}(\tilde{e}_L)\big]
    f_4^a\Big]\Big\},\label{eq_spzel}\\
\Sigma_P^a(\tilde{e}_L \tilde{e}_L) & = & \frac{g^4}{4} c_{L}\Big[
    |f^L_{\ell i}|^2 |f^L_{\ell j}|^2 \big( |\Delta^u
    (\tilde{e}_L)|^2 f_1^a-|\Delta^t (\tilde{e}_L)|^2 f_2^a
    \big) \nonumber\\
  & & \phantom{\frac{g^4}{4} c_{L}\Big[}
    + Re\{ (f^{L*}_{\ell i})^2 (f^{L}_{\ell j})^2
    \Delta^{u}(\tilde{e}_L)\Delta^{t*}(\tilde{e}_L) (f_3^a+f_4^a)\}\Big].
    \label{eq_spelel}
\end{eqnarray}
$\Sigma^a_P(Z\tilde{e}_R),\Sigma^a_P(\tilde{e}_R \tilde{e}_R)$ are
obtained by the substitutions Eq.(\ref{eq_psub}) and 
\begin{equation} \label{eq_spsubs}
f_{1,2,3}^a\to -f^a_{1,2,3} \quad \mbox{and} \quad f^a_4\to
f^a_4
\end{equation}
in Eqs.~(\ref{eq_spzel}), (\ref{eq_spelel}), where the kinematical
factors in invariant form are given by
\begin{eqnarray}
f^a_1&=& m_i (p_2 p_4)(p_1 s^a),\label{eq_kinsp1}\\
f^a_2&=& m_i (p_1 p_4)(p_2 s^a),\label{eq_kinsp2}\\
f^a_3&=& m_j [(p_1 p_3)(p_2 s^a)-(p_2 p_3)(p_1 s^a)],
\label{eq_kinsp3}\\
f^a_4 & = & i m_j \epsilon_{\mu \nu \rho \sigma} 
 p_2^{\mu} p_1^{\nu} s^{a \rho} p_3^{\sigma}.\label{eq_kinsp4}
\end{eqnarray}
The polarisation vectors $s^a(\tilde{\chi}^0_i)$ of the neutralino
$\tilde{\chi}^0_i$ are given in the laboratory system by
Eqs.~(\ref{eq_21}) -- (\ref{eq_23}).

\subsection{Leptonic 3-body decay}

Here we give the analytical expressions for the
different contributions to the
decay density matrix
for the three-body decay $\tilde{\chi}^0_i(p_3) \to
\tilde{\chi}^0_k(p_5) \ell^+(p_6) \ell^-(p_7)$, where we sum over
the spins of the final-state particles 
\cite{Moortgat-Pick:1999di,Moortgat-Pick:1999wr}.
The contributions independent of the polarisation of the neutralino
$\tilde{\chi}^0_i$ are
\begin{equation}
D(\tilde{\chi}^0_i)=D(Z Z)+ D(Z \tilde{\ell}_L)+ D(Z
\tilde{\ell}_R)+ D(\tilde{\ell}_L \tilde{\ell}_L)+
D(\tilde{\ell}_R \tilde{\ell}_R) \label{eq_sumz}
\end{equation}
with
\begin{eqnarray} D(Z Z)&=& 8
  \frac{g^4}{\cos^4\Theta_W} |\Delta^{s_i}(Z)|^2
  (L_{\ell}^2+R_{\ell}^2) \nonumber\\
 & & \Big[ |O^{''L}_{ki}|^2 (g_1+g_2) +
  [(Re O^{''L}_{ki})^2 -(Im O^{''L}_{ki})^2] g_3  \Big],
  \label{eq_dzz}\\
D(Z \tilde{\ell}_L)&=&4 \frac{g^4}{\cos^2\Theta_W} L_{\ell}
  Re\Big\{\Delta^{s_i}(Z) \Big[f^L_{\ell i} f^{L*}_{\ell k}
  \Delta^{t_i*}(\tilde{\ell}_L)
  (2O^{''L}_{ki} g_1 +O^{''L*}_{ki} g_3) \nonumber\\
 & &\phantom{4 \frac{g^4}{\cos^2\Theta_W} L_{\ell}
     Re\Big\{\Delta^{s_i}(Z) \Big[}  
  +f^{L*}_{\ell i} f^{L}_{\ell k} \Delta^{u_i*}(\tilde{\ell}_L)
  (2O^{''L*}_{ki} g_2 +O^{''L}_{ki} g_3)
  \Big]\Big\},\label{eq_dzel}\\
D(\tilde{\ell}_L \tilde{\ell}_L)&=& 2 g^4 \Big[ |f^{L}_{\ell i}|^2
  |f^L_{\ell k}|^2 \big(|\Delta^{t_i}(\tilde{\ell}_L)|^2 g_1
  +|\Delta^{u_i}(\tilde{\ell}_L)|^2 g_2\big)\nonumber\\
 & &\phantom{2 g^4 \Big[} +Re\big\{(f^{L*}_{\ell i})^2 (f^L_{\ell k})^2
  \Delta^{t_i}(\tilde{\ell}_L) \Delta^{u_i*}(\tilde{\ell}_L)\big\}
  g_3 \Big].  \label{eq_delel}
\end{eqnarray}
The quantities $D(Z\tilde{\ell}_R), D(\tilde{\ell}_R \tilde{\ell}_R)$
can be derived from Eqs.~(\ref{eq_dzel}), (\ref{eq_delel}) by
the substitutions
\begin{equation} \label{eq_substdecayP}
 L_{\ell}\to R_{\ell}, \quad
 \Delta^{t_i,u_i}(\tilde{\ell}_L)\to \Delta^{t_i,u_i}(\tilde{\ell}_R),\quad
 O^{''L}_{ki}\to O^{''R}_{ki}, \quad
 f_{\ell i,k}^L\to f_{\ell i,k}^R.
\end{equation}
The kinematical factors are
\begin{eqnarray}
g_1&=&(p_5 p_7)(p_3p_6),\label{eq_dkin1}\\
g_2&=& (p_5 p_6)(p_3 p_7),\label{eq_dkin2}\\
g_3 &=& m_i m_k (p_6 p_7).\label{eq_dkin3}
\end{eqnarray}

The contributions which depend on the polarisation of the decaying
neutralino $\tilde{\chi}^0_i$ are
\begin{equation}
\Sigma_D^a(\tilde{\chi}^0_i)= \Sigma_D^a(ZZ) +\Sigma_D^a(Z
\tilde{\ell}_L) +\Sigma_D^a(Z \tilde{\ell}_R)
+\Sigma_D^a(\tilde{\ell}_L \tilde{\ell}_L)
+\Sigma_D^a(\tilde{\ell}_R \tilde{\ell}_R) \label{eq_dssum}
\end{equation}
with\footnote{In Eqs.~(\ref{eq_dszz}) and (\ref{eq_dszel})
  misprints in Eqs.~(79) and (80) of
  \cite{Moortgat-Pick:1999di} are corrected.} 
\begin{eqnarray}
\Sigma_D^a(ZZ)&=& 8 \frac{g^4}{\cos^4\Theta_W} |\Delta^{s_i}(Z)|^2
  (R^2_{\ell}-L_{\ell}^2) \nonumber\\
 & &\times\Big[ -[(Re O^{''L}_{ki})^2 -(Im O^{''L}_{ki})^2]g^a_3+
  |O^{''L}_{ki}|^2(g^a_1-g^a_2) \nonumber\\
 & &\phantom{\times\Big[} - 2 Re(O^{''L}_{ki}) Im(O^{''L}_{ki})
  i g_4^a \Big],\label{eq_dszz}\\
\Sigma_D^a(Z \tilde{\ell}_L)&=& \frac{4
  g^4}{\cos^2\Theta_W}L_{\ell} Re\Big\{\Delta^{s_i}(Z)
  \Big[ f^L_{\ell i} f^{L*}_{\ell k} \Delta^{t_i*}(\tilde{\ell}_L)
  \big(-2 O^{''L}_{ki} g^a_1 +O^{''L*}_{ki} (g^a_3-g^a_4)\big)\nonumber\\
 & & \phantom{\frac{4 g^4}{\cos^2\Theta_W}L_{\ell} Re\Big\{}
  + f^{L*}_{\ell i} f^{L}_{\ell k} \Delta^{u_i*}(\tilde{\ell}_L)
  \big(2 O^{''L*}_{ki} g^a_2 +O^{''L}_{ki} (g^a_3+g^a_4)\big)
  \Big]\Big\},\label{eq_dszel}\\
\Sigma_D^a(\tilde{\ell}_L \tilde{\ell}_L) & = & 2 g^4 \Big[
  |f^{L}_{\ell i}|^2 |f^L_{\ell k}|^2
  [|\Delta^{u_i}(\tilde{\ell}_L)|^2 g_2^a
  -|\Delta^{t_i}(\tilde{\ell}_L)|^2 g_1^a]\nonumber\\
 & & \phantom{2 g^4 \Big[}
  + Re\big\{ (f^{L*}_{\ell i})^2 (f^L_{\ell k})^2
  \Delta^{t_i}(\tilde{\ell}_L)
  \Delta^{u_i*}(\tilde{\ell}_L)(g_3^a+g_4^a)\big\}\Big].\label{eq_dselel}
\end{eqnarray}
The contributions $\Sigma^a_D(Z\tilde{\ell}_R),
\Sigma^a_D(\tilde{\ell}_R \tilde{\ell}_R)$ are obtained from
Eqs.~(\ref{eq_dszel}), (\ref{eq_dselel}) by applying the substitutions
in Eq.~(\ref{eq_substdecayP}) and in addition
\begin{equation}
 g_{1,2,3}^a\to -g_{1,2,3}^a, \quad g_4^a \to g_4^a.
\label{eq_dssub3}
\end{equation}
The kinematical factors are
\begin{eqnarray}
g^a_1&=& m_i (p_5 p_7) (p_6 s^a), \label{eq422_3a}\\
g^a_2&=& m_i (p_5 p_6) (p_7 s^a), \label{eq_dssub4}\\
g^a_3&=& m_k [(p_3 p_6) (p_7 s^a)-(p_3 p_7) (p_6 s^a)],
\label{eq_dssub5}\\
g^a_4 & = & i m_k \epsilon_{\mu \nu \rho \sigma}
 s^{a \mu} p_3^{\nu} p_7^{\rho} p_6^{\sigma}. \label{eq_dssub6}
\end{eqnarray}

\end{appendix}


\end{document}